\documentclass[aps,prb,reprint,superscriptaddress,longbibliography]{revtex4-2}

\usepackage{graphicx}
\usepackage{latexsym}
\usepackage{amsmath}
\usepackage{amssymb}
\usepackage{amsfonts}
\usepackage{color}
\usepackage{bm}% bold math
\usepackage{verbatim}
\usepackage{hyperref}

\definecolor{MS-color}{RGB}{128,0,128}

\usepackage{framed}
\definecolor{shadecolor}{RGB}{222,222,221}
\usepackage{ulem}% for crossing out words

\begin{document}

\title{Supercurrent-induced long-range triplet correlations and controllable Josephson effect in superconductor/ferromagnet hybrids with extrinsic SOC}
%DOS and spin-resolved DOS in spin-textured S/F bilayers

\author{A. A. Mazanik}
\affiliation{Moscow Institute of Physics and Technology, Dolgoprudny, 141700 Russia}
\affiliation{BLTP, Joint Institute for Nuclear Research, Dubna, Moscow Region, 141980, Russia}

\author{I. V. Bobkova}
\affiliation{Institute of Solid State Physics, Chernogolovka, Moscow
  reg., 142432 Russia}
\affiliation{Moscow Institute of Physics and Technology, Dolgoprudny, 141700 Russia}
\affiliation{National Research University Higher School of Economics, Moscow, 101000 Russia}

\date{\today}

 %%% abstract

\begin{abstract}

We predict that long-range triplet correlations (LRTC) in superconductor/ferromagnet (S/F)  hybrids with extrinsic impurity spin-orbit coupling (SOC) can be generated and manipulated by supercurrent flowing in the superconducting leads along the S/F interfaces. The LRTC appear via two basic mechanisms. The essence of the first one is the generation of triplets by the superconducting spin Hall effect. These pairs are long-range in the ferromagnet under the appropriate mutual orientation of the condensate momentum and the ferromagnet magnetization. The second mechanism is based on the singlet-triplet conversion at the S/F interface followed by the rotation of the spin of the obtained short-range opposite-spin pairs via the spin current swapping mechanism. The structure of the supercurrent-induced LRTC is studied both for S/F bilayers and S/F/S Josephson junctions. We demonstrate that in S/F/S junctions, where the Josephson coupling is realized via the supercurrent-induced LRTC, the ground state phase can be switched between $0$ and $\pi$ in a controllable manner. The switching is performed by reversing the supercurrent  in one of superconducting leads, thus realizing a new physical principle of the $0-\pi$ shifter. 

\end{abstract}

 %%% PACS numbers
\maketitle

\section{Introduction}

Josephson junctions (JJs) are the cornerstone
elements of superconducting electronics. One of actively developing directions is an intensive search of technologies and physical 
principles allowing for the construction of superconducting transistors based on the JJs with controllable
switching between superconducting and resistive states \cite{Clark1980}. Also implementation of such structures as
$\pi$-phase shifters is of particular interest for superconducting and quantum electronics \cite{Yamashita2005,Feofanov2010,Shcherbakova2015},
which has already been demonstrated with S/F/S Josephson junctions incorporated in superconducting logical schemes  and in the qubit loop. The fundamental physical phenomenon underlying the $\pi$-phase shifters is the unusual Josephson effect characterized by the inverse current-phase relation $I=I_c \sin (\varphi+\pi)$, the so-called
$\pi$-state of a Josephson junction \cite{Buzdin1982,Buzdin2005}, which was observed in different systems like S/F/S JJs with a ferromagnetic interlayer \cite{Kontos2002,Ryazanov2001,Oboznov2006, bannykh2009josephson, robinson2006critical}, JJs with unconventional order parameter symmetry \cite{schulz2000design, smilde2002d, hilgenkamp2003ordering,ariando2005phase, lombardi2002intrinsic},  S/N/S JJs with non-equilibrium electron distribution in the normal layer \cite{Baselmans1999, golikova2021controllable}, JJs with semiconductor quantum dots \cite{vanDam2006,jorgensen2007critical}, JJs with quantum wells as interlayers \cite{ke2019ballistic}. 

Search for novel principles and possibilities  of external control of the amplitude and phase of the Josephson effect, including switching between superconducting and resistive states of the junction and switching between the $0$ and $\pi$ ground states, is very active now. There are different suggestions of externally controlled $0-\pi$ transitions in Josephson junctions. Among them one can notice the temperature induced $0-\pi$ transitions \cite{Ryazanov2001}, the width induced $0-\pi$ transitions \cite{Kontos2002, Oboznov2006},  $0-\pi$ transitions induced by the electrostatic gating \cite{vanDam2006, jorgensen2007critical}, spin-independent \cite{Volkov1995,Morpurgo1998,Baselmans1999,Huang2002,Yip2000,Heikkila2000,Wilheim1998,golikova2021controllable} and spin-dependent \cite{Bobkova2010,Bobkov2011} non-equilibrium quasiparticle distribution. The external control over the $0-\pi$ transition has also been realized in Josephson junctions containing a spin valve \cite{Gingrich2016} via the manipulation by the mutual orientation of the ferromagnets and it has been proposed theoretically via the manipulation by the exchange field orientation in S/F/S JJs with spin-orbit coupling (SOC) \cite{Bujnowski2019,Eskilt2019}.
$0-\pi$ transitions, which could be generated by the applied magnetic field, gating or by varying the JJ width were demonstrated in quantum wells \cite{ke2019ballistic}. Main efforts towards realization the superconducting transistor, which implies the control over the supercurrent amplitude, have been focused on the systems with Josephson currents controlled by electrostatic gates. This concept has been realized in mesoscopic systems with metallic \cite{Paolucci2019,DeSimoni2018} and semiconducting interlayers \cite{Clark1980,Larsen2015,Casparis2016,Doh2005,Abay2014}. S/F/S JJs provide additional   possibilities for external switching of the Josephson current amplitude. For example, one of suggestions was to exploit S/F/S Josephson junctions under nonequilibrium quasiparticle distribution in the weak link \cite{Bobkova2012}. The other possibility is to manipulate by the amplitude of the so-called long-range triplet correlations (LRTC). In many cases the Josephson current in S/F/S JJs via strong ferromagnets is only carried by the LRTC because they can penetrate at large distances into the ferromagnetic material  \cite{Bergeret2001,Bergeret2001_2,Kadigrobov2001,Fominov2003,Bergeret2005, Houzet2007,fominov2007josephson,Fominov2010,Halterman2007,Halterman2008,Zhu2010,Alidoust2014,Keizer2006,Eschrig2003,Eschrig2008,Robinson2010,Braude2007,Robinson2010_2,Singh2016,Khaire2010,Mironov2015,Halterman2016,Halterman2018,Srivastava2017,Linder2015,eschrig2015spin,Niu2012,Bergeret2013,Bergeret2014,Jacobsen2015,Eskilt2019,Bujnowski2019}.

Recently it has been predicted that the superconducting LRTC in S/F hybrids with Rashba-type SOC interfaces can be generated by the condensate motion along the S/F interfaces \cite{silaev2020odd,Bobkova2021}. It has been demonstrated that switching on the LRTC in the S/F/S Josephson junction by the Meissner currents induced by the applied magnetic field or by the the electromagnetic radiation allows for a realization of a controllable superconducting transistor. Here we demonstrate that the effect of the LRTC generation by the supercurrent is not  restricted by the hybrids with interface Rashba coupling and exists in a wider class of systems with external impurity SOC. We investigate the LRTC generation in S/F bilayers and S/F/S Josephson junctions. In the last case the phase difference between the LRTC induced at the opposite S/F interfaces depends on the directions of the condensate motion in the both superconducting leads. Thus we demonstrate that the supercurrent-induced LRTC in Josephson junctions provide a possibility of a supercurrent-controllable $0-\pi$ transition in the junction. Therefore, it appears that the long-range triplet superconductivity generated by the moving condensate is a very interesting phenomenon from the point of view of the superconducting electronics allowing for the total control over both the amplitude and the ground state phase of the Josephson current.

The paper is organized as follows. In Sec.~\ref{model} we describe the physical model under consideration and the formalism we use. In Sec.~\ref{LRT} the behavior of LRTC in S/F hybrids of different types is investigated. Sec.~\ref{bilayers} is devoted to the structure and characteristic features of the LRTC in S/F bilayers, while Secs.~\ref{SFS_high} and \ref{SFS_low} present results of the LRTC study in S/F/S Josephson junctions with highly-transparent and low-transparent interfaces, respectively. Sec.~\ref{controllable} describes the mechanism of $0-\pi$ switching by the supercurrent. Our conclusions are formulated in Sec.~\ref{conclusions}.

\section{Model}
\label{model}

\begin{figure}[htp]
    \centering
    \includegraphics[width=0.8\linewidth]{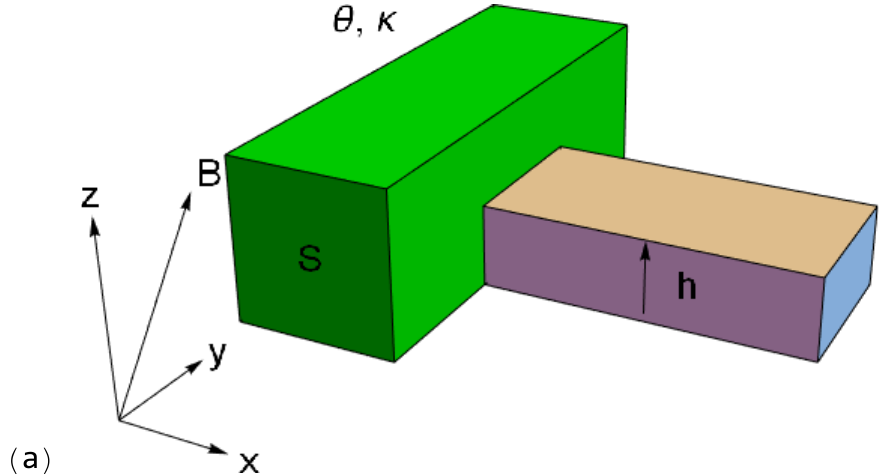}\\
    \includegraphics[width=0.8\linewidth]{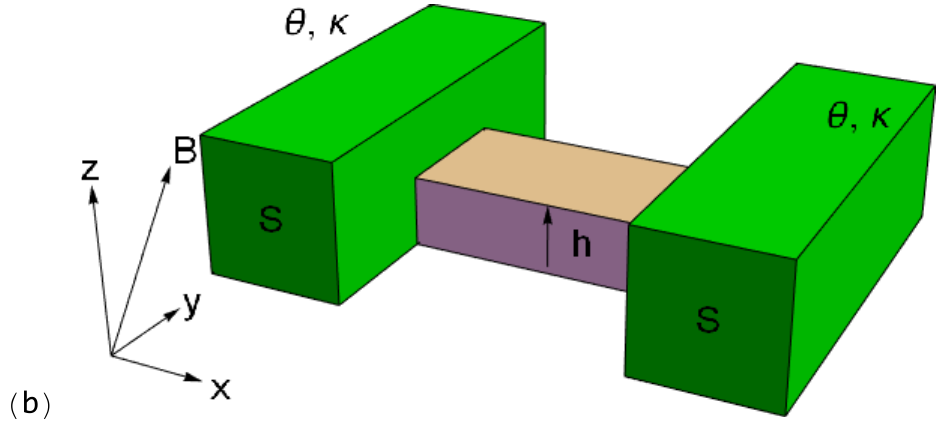}
    \caption{\label{fig:system}Sketches of the systems under consideration. (a) S/F geometry. (b) S/F/S geometry. }
    \label{fig:my_label}
\end{figure}

We consider two types of structures: a S/F bilayer and a S/F/S Josephson junction. The S/F bilayer is  shown in Fig.~\ref{fig:system}(a), which consists of an infinite $s$-wave superconducting layer $x \leq 0$ and a ferromagnetic layer $0 < x \leq d_F$ with the exchange field $\mathbf{h} = (0,0,h)$. The S/F/S junction shown in Fig.~\ref{fig:system}(b) consists of two infinite superconducting leads separated by the ferromagnetic interlayer at $0<x<d_F$ with the same exchange field. It is assumed that there is an extrinsic (impurity-induced) SOC in the superconductors. This type of SOC can be realized in different superconducting materials like $Nb$ \cite{jeon2018spin}, $NbN$ \cite{wakamura2015quasiparticle}, $V$ \cite{wang2017large} and others. Also it is known, that a dopping of non-magnetic materials with the heavy atoms like $Ir$ or $Pt$ may induce the non-zero impurity SOC \cite{niimi2011extrinsic,ramaswamy2017extrinsic}, so it is possible that such dopping may help in creation of new superconductors with the impurity SOC. 

To take into account the effect of the extrinsic SOC a quasiclassical theory of superconductivity in the diffusive limit, formulated in terms of the Usadel equation for the Green's functions, has been developed  in Refs.~\cite{bergeret2016manifestation,espedal2017spin,huang2018extrinsic, virtanen2021magnetoelectric}  and has been used for prediction of a number of SOC-related effects in superconducting structures \cite{bergeret2016manifestation,espedal2017spin}. In the framework of this theory the extrinsic SOC enters the Usadel equation via the spin Hall angle $\theta$,  the spin current swapping coefficient $\kappa$ and the spin-orbit scattering time $\tau_{so}$. Here we consider the effect of the LRTC generated by the moving condensate in the framework of this theory. The interfaces between the superconducting layers and the ferromagnetic layer are treated in both transparent and tunnel limits. The systems are placed into an external magnetic field with the vector potential $\mathbf{A} = (0, A_y(x), A_z(x))$, $\operatorname{div}{\mathbf{A}} = 0$ and $\mathbf{B} = \operatorname{rot}{\mathbf{A}} =- \hat{\mathbf{e}}_y \partial_x A_z(x) + \hat{\mathbf{e}}_z \partial_x A_y(x)$. 

In order to describe our systems we use the modified Usadel equation from \cite{virtanen2021magnetoelectric} 
\begin{equation} \label{eq:UsadelFull}
	\begin{aligned}
		\left[ \omega_n \check{\tau}_3 - i \check{\Delta} + i (\mathbf{h}\cdot\hat{\bm{\sigma}})\check{\tau}_3\text{, }\check{g}\right] + \tilde{\nabla}_k \check{J}_k =
		\\= -\frac{1}{8 \tau_{so}} \left[\hat{\sigma}_a \check{g}\hat{\sigma}_a\text{, }\check{g}\right] + \check{T}.\\
	\end{aligned}
\end{equation}
 $\omega_n = \pi T (2 n + 1)\text{, }n = 0, \pm 1, ...$ are the fermionic Matsubara frequencies at a temperature $T$. The check notation $\check{\#}$ is used for matrices in the particle-hole space, the hat notation $\hat{\#}$ is used for matrices in the spin space. $\check{\tau}_0 = \check{1}$, $\check{\tau}_{1}$, $\check{\tau}_3$, $\check{\tau}_3$ and $\hat{\sigma}_0 = \hat{1}$, $\hat{\sigma}_x$, $\hat{\sigma}_y$, $\hat{\sigma}_z$ are the Pauli matrices in the particle-hole and the spin spaces, respectively.

The matrix Green function $\check{g}$ in the particle-hole and the spin spaces can be parameterized as follows:
\begin{equation}
    \check{g} = \begin{pmatrix}
			\hat{g}, &\hat{f} \\
			-\hat{f}^c, & -\hat{g}^c
		\end{pmatrix}, 
\end{equation}
with the normalization condition  $\check{g}^2 = \check{\tau}_0$. The object $\check{g}$ contains some constrictions between its components. They could be implemented via the $\tilde{}$ conjugation \cite{eschrig2015spin}, which reads in the basis we use $\tilde{q}(\mathbf{R}, i\omega_n) = - q^\star(\mathbf{R},  i\omega_n)$. If we define $\hat{g} = g_s + \mathbf{g}_t \cdot\hat{\bm{\sigma}}$, $\hat{f} = f_s + \mathbf{f}_t \cdot\hat{\bm{\sigma}}$, then we end up with $\hat{g}^c = - \tilde{g}_s + \tilde{\mathbf{g}}_t\cdot\hat{\bm{\sigma}}$, $\hat{f}^c = \tilde{f}_s - \tilde{\mathbf{f}}_t\cdot\hat{\bm{\sigma}}$.  $f_s$ and $\bm f_t$ account for the singlet and triplet superconducting correlations, respectively. $g_s$ and $\bm g_t$ describe spin-independent and spin-dependent normal Green's functions.  

In our basis the superconducting order parameter takes the form
\begin{equation}
    \check{\Delta} = \begin{pmatrix}
			0, & \Delta \\
			-\Delta^\star, & 0\\
		\end{pmatrix}.
\end{equation}
Further we will neglect variations of the order parameter due to the inverse proximity effect, assuming $\Delta(x)=const$ in the superconducting leads.

The generalized matrix current $\check{J}_k$ and the torque $\check{T}$ entering Eq.~(\ref{eq:UsadelFull}) take the form \cite{virtanen2021magnetoelectric}: 
\begin{equation}\label{eq:GenCurrentAndTorque}
\begin{aligned}
    &\check{J}_k =  - D\check{g} \tilde{\nabla}_k \check{g} + &\\
	&+\frac{D\epsilon_{kja}}{4} \left[ \theta\left\{\tilde{\nabla}_j \check{g}\text{, } \hat{\sigma}_a\right\}\check{g} + i\kappa\left[\check{g}\tilde{\nabla}_j\check{g}\text{, }\hat{\sigma}_a \right]\check{g}\text{, }\check{g} \right],&\\
	&\check{T} = -\frac{D\theta}{4} \epsilon_{kja}\left[\check{g}\tilde{\nabla}_k \check{g} \tilde{\nabla}_j \check{g}\text{, }\hat{\sigma}_a \right] +&\\
	&+ \frac{i D\kappa}{4}\epsilon_{kja}\left[\tilde{\nabla}_k \check{g} \tilde{\nabla}_j \check{g}\text{, }\hat{\sigma}_a\right].&
\end{aligned}
\end{equation} 
$\epsilon_{kja}$ is the Levi-Civita tensor.  $D$ is the diffusion coefficient in an appropriate material.

Vector potential $\mathbf{A}(x)$ enters Eqs.~ (\ref{eq:UsadelFull}) and (\ref{eq:GenCurrentAndTorque}) in a gauge invariant manner via the gauge-covariant derivative
\begin{equation}
    \begin{aligned}
		\tilde{\nabla}_k \check{M} = \nabla_k \check{M} + i \left[\frac{\mathbf{p}_s}{2} \check{\tau}_3,\text{ } \check{M}\right].\\ 
	\end{aligned}
\end{equation} 
Here we have already recasted a non-constant  phase of the superconducting order parameter $\varphi(x)$ and the vector potential $\mathbf{A}(x)$ to the condensate momentum 
$\check{g} \to \exp{(im\varphi\check{\tau}_3/2)}\check{g}\exp{(-im\varphi\check{\tau}_3/2)}$, $\mathbf{p}_s = m\nabla \varphi - \frac{2e}{c}\mathbf{A}$.

The spin Hall angle $\theta$, the spin current swapping coefficient $\kappa$ and the spin-orbit scattering time $\tau_{so}$ are non-zero constants only in the superconductors. We will use an approximation $\tau_{so} \gg 1/T_c$  in final answers, unless otherwise is specified. The exchange field $\mathbf{h}$ exists only in the ferromagnetic layer $0 < x < d_F$. The London penetration depth is supposed to be the largest scale in the system $\lambda_L \gg (\xi_{(S,F)}, d_F)$ with the superconducting coherence lengths $\xi_S = \sqrt{\frac{D_S}{2 \pi T_c}}$, $\xi_F = \sqrt{\frac{D_F}{h}}$.

One should supplement Eq.~(\ref{eq:UsadelFull}) by the boundary conditions. If we assume that the interfaces are fully transparent, we imply continuity of the Green function $\check{g}(x)$ and the matrix current $\check{J}_x$ at the S/F interfaces \cite{huang2018extrinsic} 
\begin{equation} \label{eq:boundcoundSF}
	\check{g}_S= \check{g}_F\text{, }	\check{J}_{xS} = \check{J}_{xF}.
\end{equation}
In this case it is convenient to set for simplicity $D_S = D_F = D$, $\nu_{0S} = \nu_{0F} = \nu_0$, where $\nu_{0(S,F)}$ are the densities of states at the Fermi level in the superconductor and the ferromagnet, respectively.  

If we assume that the S/F interfaces are tunnel, we use the modified Kupriyanov-Lukichev boundary conditions proposed in Ref.~\cite{bergeret2016manifestation} 
\begin{equation} \label{eq:boundcoundTSF}
\begin{aligned}
	\mathbf{n}\cdot\check{\mathbf{J}}= \frac{D_F}{2 R_b \sigma_F} \left[\check{g}_S\text{, }\check{g}_F\right],\\
	\frac{\sigma_S}{D_S} \check{J}_{xS} = \frac{\sigma_F}{D_F}\check{J}_{xF}.
\end{aligned}
\end{equation}
$\mathbf{n} = (\pm1, 0, 0)$ is a vector perpendicular to the interface, $\check{\mathbf{J}} = (\check{J}_x, \check{J}_y, \check{J}_z)$. $R_b$ is the barrier resistance per unit cross-section of the junction, $\sigma_S$ and $\sigma_F$ are the conductivities of the superconducting and ferromagnetic regions respectively. At the external surface of the S/F bilayer $x=d_F$ we imply $\check{J}_{xF} = 0$.

\section{Supercurrent-induced generation of LRT superconductivity in S/F hybrids}

\label{LRT}

First of all, we are going to discuss qualitative physics of the LRTC generation by supercurrent in S/F hybrids with impurity-induced SOC. The key parameters here are the spin Hall angle $\theta$ and the spin current swapping coefficient $\kappa$. In the presence of nonzero condensate momentum $\bm p_s$ each of them generates the LRTC according to its own physical mechanism, which are physically different. 

The spin Hall angle accounts for an analog of the well-known spin Hall effect \cite{Sinova2015} in superconducting systems. Namely, it has been demonstrated \cite{bergeret2016manifestation} that in the presence of $\theta \neq 0$ the finite momentum of the condensate induces triplet components of the condensate at the edges of the finite-width S-electrode. Vector structure of the triplet pair wave function is constructed via the edge normal $\bm n$ and the condensate momentum $\bm p_s$ as $\bm f_t \propto \bm p_s \times \bm n$.  For the case under consideration the role of the edges is played by the S/F interface. Upon entering the ferromagnetic region the triplet pairs feel the ferromagnet exchange field $\bm h$. Let us choose the spin quantization axis along $\bm h$. Then if $\bm h$ is aligned with $\bm p_s \times \bm n$ the vector wave function of the triplet correlations $\bm f_t = f_t \bm e_z$ has the only nonzero $z$-component. That is, the triplet correlations consist of opposite-spin triplet pairs, which decay in the ferromagnet at the distance of the magnetic coherence length $\xi_F$ \cite{Buzdin2005}. It is a very short distance of the order a few nanometers in conventional ferromagnets like $Fe$ or permalloy. At the same time, if $\bm h$ has a nonzero component perpendicular to $\bm p_s \times \bm n$ [what means that only $p_{sz}$ component of the condensate momentum matters], the vector of triplet correlations $\bm f_t$ is no longer along the $z$-axis and has a perpendicular component, which corresponds to equal-spin pairs. It is long-range in ferromagnets \cite{Bergeret2005} and decays at the length scale of the normal state coherence length $\propto\sqrt{D_F/2 \pi T_c}$, which is of the order of tens to hundreds nanometers depending on the material. Therefore, a proper choice of the condensate momentum direction allows to make the triplet pairs, which are generated via the superconducting spin Hall effect, long-range.

The second mechanism of the LRTC generation is realized via the spin current swapping coefficient $\kappa$. It is completely different. In contrast to the previous mechanism the triplet pairs are not generated in the system in the absence of the exchange field via this mechanism, at least, in the framework of the quasiclassical formalism used here. This statement is supported further by direct calculations. Now the two stages of the LRTC generation are the following. At first the short-range opposite-spin triplet pairs are generated at the S/F interface via the conventional singlet-triplet conversion \cite{Buzdin2005}. Then the spin of these triplet pairs, which have nonzero total momentum due to the condensate motion along the interface, is partially rotated via the spin current swapping mechanism \cite{Lifshits2009swapping}. In order to create LRT pairs via this mechanism, that is, in order to generate $\bm f_t$ components perpendicular to $\bm h = h \bm e_z$, the condensate momentum again should have a nonzero component along $z$-direction. 

Further we consider several specific examples of S/F hybrids, where the LRTC are generated according to these general mechanisms.

\subsection{S/F bilayer with absolutely transparent interfaces}
\label{bilayers}

In this section we consider $S/F$ bilayer shown in Fig.~\ref{fig:system}(a). We perform a linearization of Eq.~(\ref{eq:UsadelFull}) with respect to the anomalous Green's function in the regime $T \to T_c$, $\Delta \to 0$. Then the Green's function takes the form $\check{g} = \check{\tau}_3 \operatorname{sgn}{\omega_n} + \check{f}$. 
One can see that the torque $\check{T}$ in Eq.~(\ref{eq:GenCurrentAndTorque}) is  of the second order with respect to $\hat{f}$, $\hat{f}^c$, therefore it does not enter the linearized Usadel equation. The terms $\epsilon_{kja} \tilde{\nabla}_k \check{g} \tilde{\nabla}_j \check{g} \propto \epsilon_{kja} \tilde{\nabla}_k \check{f} \tilde{\nabla}_j \check{f} $ in $\tilde{\nabla}_k\check{J}_k$ are also of the second order with respect to the anomalous Green's function and are neglected. The term $D \epsilon_{kja}\tilde{\nabla}_k \tilde{\nabla}_j\check{g} \propto D e\vert\operatorname{rot}{\mathbf{A}} \vert /c  \propto h (\xi_F p_s)(\xi_F/\lambda_L) \ll h $ in the ferromagnet or $\propto T_c (\xi_S p_s)(\xi_S/\lambda_L) \ll T_c$ in the superconductor. Therefore, it is also neglected. 
In the superconducting region the resulting linearized equation for $\hat{f} = f_s + \mathbf{f}_t\cdot\hat{\bm{\sigma}}$ takes the form:
\begin{equation} \label{eq:UsadelLinInS}
\begin{cases}
	\vert\omega_n\vert f_s + i\Delta - \frac{D}{2}\left(f''_s - p^2_s f_s\right) = 0,\\
	\vert\omega_n\vert \mathbf{f}_t - \frac{D}{2}\left(\mathbf{f}''_t -p^2_s \mathbf{f}_t\right) = -\frac{1}{2\tau_{so}}\mathbf{f}_t.  
\end{cases}
\end{equation}
Here $p^2_s = p^2_{sy} + p^2_{sz}$. The condensate momentum changes on a  scale, which is determined by the London penetration depth $\lambda_L$. This depth is assumed to be the largest scale in the system. Therefore, the condensate momentum can be treated locally as a constant quantity in the superconducting leads. However, its spatial dependence can be essential  in the ferromagnetic region, for example, in the Josephson junction setup (see Eq.~(\ref{eq:psInJJ}) in the Appendix). However, here we assume $p_s\xi_S \ll 1$, what allows for treating only linear terms in $p_s$ and neglecting all the orbital depairing effects. At $x \leq 0$ solution of Eq.~(\ref{eq:UsadelLinInS}) takes the form: 
\begin{align}
\label{eq:UsadelSSol}
f_s(x) &= -\frac{i \Delta}{|\omega_n|}+ \Bigl( \frac{i \Delta}{|\omega_n|} + f_{s0} \Bigr) e^{\lambda_{Ss}x}, \nonumber \\
\bm f_t (x) &= \bm f_{t0} e^{\lambda_{St}x},
\end{align}
where $f_{s0} \equiv f_s(x=0)$ and $\bm f_{t0} = (f_{x0},f_{y0},f_{z0})^T \equiv \bm f_t(x=0)$ are the values of the singlet and triplet anomalous Green's function at the S/F interface, $\lambda_{Ss} = \sqrt{\frac{2\vert\omega_n\vert}{D} + p^2_s} \approx \sqrt{\frac{2 \vert \omega_n \vert}{D}}$, $\lambda_{St} = \sqrt{\frac{2\vert\omega_n\vert}{D} + p^2_s + \frac{1}{D \tau_{so}}} \approx \sqrt{\frac{2\vert\omega_n\vert}{D}  + \frac{1}{D \tau_{so}}} $.

In the ferromagnetic layer we obtain
\begin{equation}  \label{eq:UsadelLinInF}
\begin{cases}
	\vert\omega_n\vert f_s +i\operatorname{sgn}{\omega_n}(\mathbf{h}\cdot\mathbf{f}_t) - \frac{D}{2} f''_s \iffalse \left(f''_s - p^2_s f_s\right) \fi = 0,\\
	\vert\omega_n\vert \mathbf{f}_t + i\operatorname{sgn}{\omega_n}\mathbf{h}f_s - \frac{D}{2} \mathbf{f}''_t \iffalse \left(\mathbf{f}''_t -p^2_s \mathbf{f}_t\right) \fi = 0.  	
\end{cases}
\end{equation}
At $x < 0 \leq d_F$ the solution of Eqs.~(\ref{eq:UsadelLinInF}) takes the form:
\begin{align}
\label{eq:UsadelFSol}  
\begin{pmatrix}
		f_s\\
		f_{z}
	\end{pmatrix} &=  \begin{pmatrix}
		\frac{f_{s0} + f_{z0}}{2}\\
		\frac{f_{s0} + f_{z0}}{2}\\
	\end{pmatrix}\frac{\cosh{\left[\lambda_{F+}(x-d_F)\right]}}{\cosh{\left[\lambda_{F+}d_F\right]}} + \nonumber \\ 
	&+
	\begin{pmatrix}
		\frac{f_{s0} - f_{z0}}{2}\\
		-\frac{f_{s0} - f_{z0}}{2}\\
	\end{pmatrix}\frac{\cosh{\left[\lambda_{F-}(x-d_F)\right]}}{\cosh{\left[\lambda_{F-}d_F\right]}}, \nonumber \\
	\begin{pmatrix}
		f_{x}\\
		f_{y}
	\end{pmatrix} &=  
	\begin{pmatrix}
		f_{x0}\\
		f_{y0}
	\end{pmatrix} \frac{\cosh{\left[\lambda_{FL}(x-d_F)\right]}}{\cosh{\left[\lambda_{FL}d_F\right]}}.
\end{align}
Here  $\lambda_{FL} = \sqrt{\frac{2\vert\omega_n\vert}{D}} $, $\lambda_{F\pm} = \sqrt{\frac{2 (\vert\omega_n\vert \pm ih\operatorname{sgn}{\omega_n})}{D} } \approx (1 \pm i\operatorname{sgn}{\omega_n})\sqrt{\frac{h}{D}}$ as $\xi_F = \sqrt{\frac{D}{h}} \ll \xi_S$. The boundary conditions $\check{g}(x = -0) = \check{g}(x = +0)$ and $\check{J}_x(x = d_F) = 0$ have already been implemented. 

Implementing the second boundary condition in Eq.~(\ref{eq:boundcoundSF}) [see details of the calculation in the Appendix] we obtain the values of the singlet and triplet anomalous Green's functions at the S/F interface. The general expression are rather cumbersome and we do not write them down here.  However, a simplified analytical treatment is possible under the reasonable assumption $\tau_{so} \gg 1/T_c$. In that case $\lambda_{Ss} = \lambda_{St} = \lambda_{FL} = \lambda_s$. We also assume $d_F \gg (\xi_F\text{, }\xi_S)$, what results in $\tanh{\lambda_{(F\pm\text{, }FL)} d_F} \approx 1$. Up to the  first  order in $\theta$, $\kappa$  and up to the first order in $ T_c/h$ we obtain
\begin{eqnarray}
\begin{pmatrix}
		f_{s0}\\
		f_{z0}
	\end{pmatrix} = \begin{pmatrix}
		f_{s0}\\
		f_{z0}
	\end{pmatrix}^{(0)} +\begin{pmatrix}
		f_{s0}\\
		f_{z0}
	\end{pmatrix}^{(1)},  \nonumber \\
\begin{pmatrix}
		f_{s0}\\
		f_{z0}
	\end{pmatrix}^{(0)} = -\frac{\Delta}{\vert\omega_n\vert \iffalse + \frac{D p^2_s}{2} \fi}\frac{i\lambda_s \xi_F}{Z}  \begin{pmatrix}
		 1 + \lambda_s \xi_F \\
		-i\operatorname{sgn}{\omega_n}  \\
	\end{pmatrix}, \nonumber \\
\begin{pmatrix}
		f_{s0}\\
		f_{z0}
	\end{pmatrix}^{(1)} = \frac{\theta p_{sy} \xi_F}{Z}
	\hat K \begin{pmatrix}
		f_{s0}\\
		f_{z0}
	\end{pmatrix}^{(0)},
\label{eq:FSolLinInThetaKappa_sz}
\end{eqnarray}
where $Z=1 + \left(1 + \lambda_s \xi_F \right)^2$ and $\hat K = \operatorname{diag} \{ 2\text{, } -\lambda_s\xi_F (2+ \lambda_s \xi_F) \}$. Superscripts $(0)$ and $(1)$ denote zero and first order terms with respect to $\theta p_s \xi_S$ and  $\kappa p_s \xi_S$, respectively. $f_{x0}$ and $f_{y0}$ are of the first order with respect to these parameters and take the form:
\begin{eqnarray}
\begin{pmatrix}
		f_{x0}\\
		f_{y0}
	\end{pmatrix} =  \frac{\Delta}{\vert\omega_n\vert \iffalse + \frac{D p^2_s}{2} \fi}\frac{i\operatorname{sgn}{\omega_n} p_{sz}\xi_F}{2Z}  \begin{pmatrix}
		 \kappa \\
		i \theta (1+\lambda_s \xi_F)  \\
	\end{pmatrix}. 
\label{eq:FSolLinInThetaKappa_xy}
\end{eqnarray}

From Eq.~(\ref{eq:FSolLinInThetaKappa_xy}) we see that $f_{x}$ and $f_{y}$ are produced by the condensate motion. Moreover, only $p_{sz}$ component of the condensate momentum plays role in the LRTC generation, as it is qualitatively discussed above. According to Eq.~(\ref{eq:UsadelFSol}) these correlations decay on the characteristic scale $\lambda_s^{-1}$, which is large in comparison with the decay scale $[\operatorname{Re}\lambda_{F\pm}]^{-1}$ of the short-range correlations (SRC). 

\begin{figure}[htp]
    \centering
    \includegraphics[width=0.95\linewidth]{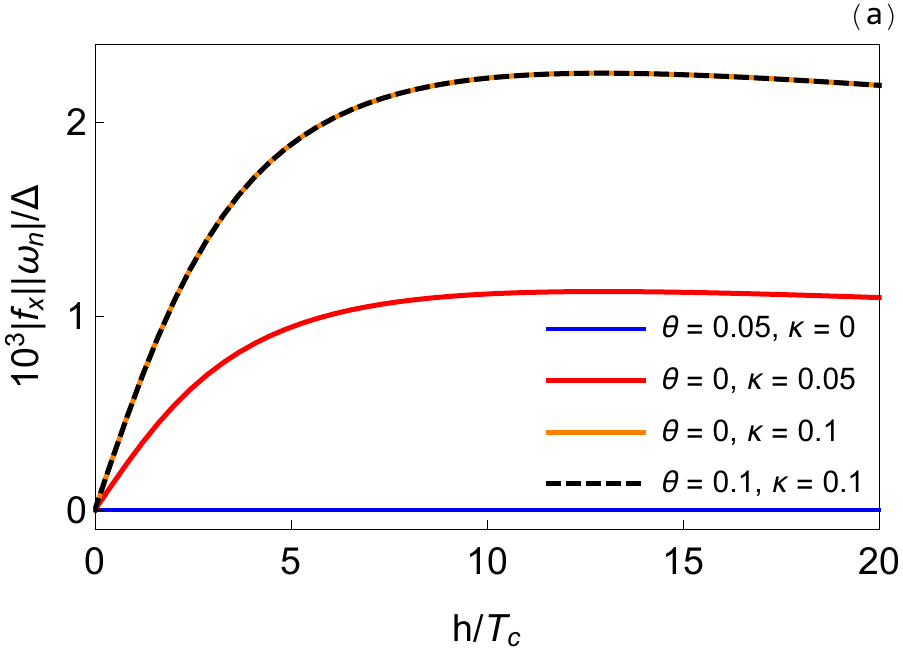}\\
    \includegraphics[width=0.95\linewidth]{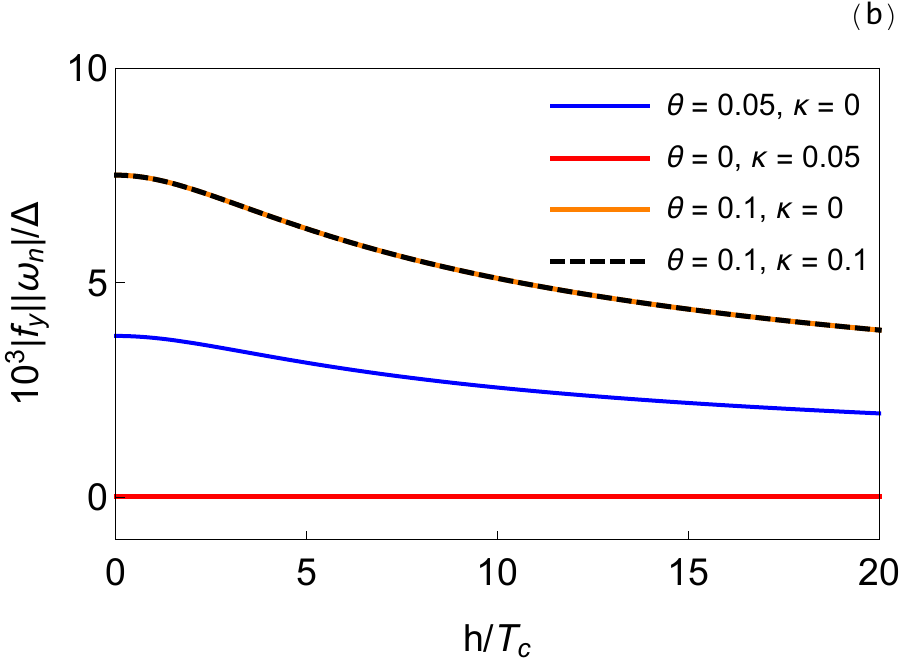}
    \caption{LRTC $\vert f_x(h) \vert$ (a) and $\vert f_y(h) \vert$ (b) as functions of the exchange field. In both panels dashed black and solid orange lines coincide. The parameters are $D_S = D_F$, $\omega_n = \pi T_c$, $p_{sy} \xi_S = 0$, $p_{sz} \xi_S = 0.3$.  }
    \label{fig:f_h}
\end{figure}

The behavior of the LRTC at arbitrary $\bm h$ values is shown in Fig.~\ref{fig:f_h}. Eq.~(\ref{eq:FSolLinInThetaKappa_xy}) corresponds to the limit $h/T_c \gg 1$ in this figure. Data presented in Fig.~\ref{fig:f_h} support the physical picture of the LRTC generation, described above. Indeed, at $\theta \neq 0$, $\kappa = 0$ the triplet correlations survive at $h \to 0$. It can be  directly checked that at $h=0$: 
\begin{equation}
    \bm f_{t0} = \frac{\operatorname{sgn}{\omega_n} \Delta \theta}{4\vert \omega_n \vert \lambda_s}[\bm n \times \bm p_s].
    \label{f_h0}
\end{equation}
These condensate-induced triplet correlations in the absence of the exchange field  have already been reported in the literature \cite{bergeret2016manifestation} and are understood as a superconducting manifestation of the spin Hall effect. $f_{y} \propto p_{sz}$ component of triplet correlations Eq.~(\ref{f_h0}) is long-range in the ferromagnet. In the case $\theta = 0$, $\kappa \neq 0$ $f_{x,y}$ vanishes at $h \to 0$ thus supporting our qualitative statement that the mechanism of LRTC generation via $\kappa$ requires preformed short-range triplets produced by the triplet-singlet conversion at the S/F interface. It is also seen from Fig.~\ref{fig:f_h} that $\theta$ does not influence $f_x$ component of the LRTC and $\kappa$ does not contribute to $f_y$ in the whole range of $h$, not only in the limit of $h/T_c \gg 1$, as it is suggested by Eq.~(\ref{eq:FSolLinInThetaKappa_xy}).

Beyond the linear approximation with respect to $p_s \xi_S$ there is an optimal condensate momentum corresponding to the maximal amplitude of the long-range triplet correlations because the amplitude  of the LRTC is bounded  from above  by the depairing effects $\propto p_s^2$, which have been neglected in the framework of our linear approximation. 

\subsection{S/F/S Josephson junction with absolutely transparent interfaces}
\label{SFS_high}

In this section we consider S/F/S Josephson junction shown in Fig.~\ref{fig:system}(b), which is composed of  $s$-wave superconducting leads $x \leq 0$ and $x \geq d_F$ with the extrinsic spin-orbit coupling and the ferromagnetic interlayer $0 < x < d_F$ with the exchange field $\mathbf{h} = (0,0,h)$. It is assumed that $d_F \gg \xi_F$. The parameters of the leads $\theta$, $\kappa$, $\vert \Delta \vert$, $\tau_{so}$ are the same as in the previous section.  We also assume the phase difference $\varphi$ between the leads, that is $\Delta(x \leq 0) = \vert\Delta\vert e^{i\varphi/2}$, $\Delta(x \geq d_F) = \vert\Delta\vert e^{-i\varphi/2}$.  The $S/F$ interfaces are fully transparent, and the boundary conditions at the right S/F interface now read  $\check{J}_x(x = d_F-0) = \check{J}_x(x = d_F+0)$ and $\check{g}(x = d_F-0) = \check{g}(x = d_F + 0)$ instead of $\check{J}_x(x = d_F) = 0$. The external magnetic field is again in the $zy$ plane  $\mathbf{A} = (0, A_y(x), A_z(x))$,  $\mathbf{B} = \operatorname{rot}{\mathbf{A}} =- \hat{\mathbf{e}}_y \partial_x A_z(x) + \hat{\mathbf{e}}_z \partial_x A_y(x)$  as in the previous section.

Under the condition $d_F \gg \xi_F$ the Josephson current is only carried by the LRTC. The SRC $f_s$ and $f_{tz}$, generated at the both S/F interfaces, do not overlap and, therefore, coincide with that ones calculated in the previous section with the substitution of the appropriate $\Delta(x)$.   We also should take into account that the condensate momentum, generated by the superconductors in response to the external magnetic field, has opposite directions in the leads, as it is demonstrated in the Appendix. Then we find the following answer for the LRTC in the $F$ interlayer (see details of the derivation in the Appendix): 
\begin{equation} \label{eq:LRTCideal}
\begin{aligned}
	\begin{pmatrix}
		f_{x}\\
		f_{y}
	\end{pmatrix} =\frac{i\vert\Delta\vert p_{sz}\xi_F e^{-\lambda_s d_F/2}}{\omega_n Z} \times \\ \cos\bigl[  \frac{\varphi}{2} + i\lambda_s(x-\frac{d_F}{2})\bigr]\begin{pmatrix}
		\kappa \\	
		i\theta (1 + \lambda_s \xi_F)
	\end{pmatrix}  
\end{aligned}
\end{equation} 

Now we are ready to calculate the Josephson current. It reads as
\begin{equation} \label{eq:JJCDideal}
\begin{aligned}
	j_x(\varphi) = i \frac{\sigma \pi T_c}{e} \sum_{\omega>0} \left[ \tilde{f}_{x} \partial_x f_{x} + \tilde{f}_{y} \partial_x f_{y} - \operatorname{c.c.} \right] =
	\\ =-\frac{4 \sigma }{e} \frac{\vert \Delta\vert^2}{\pi T_c} \iffalse \frac{1}{\left[1 + \frac{D p^2_s}{2 \pi T_c}\right]^2} \fi \frac{\kappa^2 + \theta^2\left(1 + \frac{\xi_F}{\xi_S}\right)^2 }{\left[1 + \left(1 + \frac{\xi_F}{\xi_S}\right)^2\right]^2}  \frac{ (\xi_F p_{sz})^2}{4\xi_S} e^{-d_F/\xi_S} \sin{\varphi}.
\end{aligned}
\end{equation}
Here $\sigma = 2e^2 \nu_0 D$ is the Drude conductivity of the ferromagnetic region. In the bottom line of Eq.~(\ref{eq:JJCDideal}) only the lowest Matsubara frequency is taken into account. In the limit $\frac{h}{T_c} \gg 1$ \iffalse and weak external magnetic field $p_s \xi \ll 1$ \fi we have
\begin{equation} \label{eq:CPRideal}
\begin{aligned}
	j_x(\varphi) = -\frac{ \sigma}{e} \frac{\vert\Delta\vert^2}{\pi T_c} e^{- d_F/\xi_S} (\kappa^2 + \theta^2) \frac{ (p_{sz} \xi_F)^2}{4\xi_S}  \sin{\varphi} =\\
	= -\frac{j_{c0}}{16}  (\kappa^2 + \theta^2)  (p_{sz} \xi_S)^2  \frac{\xi^2_F}{\xi^2_S}
	  \sin{\varphi} =  - j_{c} \sin{\varphi}.
\end{aligned}
\end{equation}
Here $j_{c0} =\frac{4\sigma}{e} \frac{\vert\Delta\vert^2}{\pi T_c} e^{-d_F/\xi_S} \frac{1}{\xi_S}$ is the critical current density of a corresponding S/N/S junction \cite{golubov2004current}. For estimates of the critical current amplitude we take typical  parameters of the S/F/S JJs \cite{fominov2007josephson}: the junction area is $ S\sim 50 \text{ }\mu m\times50\text{ }\mu m$, the conductivity is $\sigma \sim (50\text{ }\mu\Omega\text{ }cm)^{-1}$, the width of the ferromagnetic layer is $d_F \sim 5\xi_F$, the diffusion coefficient is $D \sim 10\text{ }cm^2/s$, the exchange field is $h \sim 500\text{ }K$, so $\xi_F \sim 3\text{ }nm$, the typical order parameter is $\vert\Delta\vert \sim 1 \text{ }K$ ($T \to T_c$ regime), the critical temperature is $T_c \sim 10\text{ }K$, so $\xi_S \sim 30 \text{ }nm$, the spin Hall angle is $\theta\sim 0.1 - 0.001$  \cite{jeon2018spin,wang2017large,wakamura2015quasiparticle,hoffmann2013spin},  the swapping coefficient is $\kappa \sim 0.1 - 0.001$ \cite{Lifshits2009swapping}. With these data we obtain $\frac{I_c}{I_{c0}} \sim 10^{-10} - 10^{-6}$, for $p_{sz}\xi_S \sim 0.3$ and $I_c = j_c S$. For $I_{c0} \sim 10^{0} \text{ }A$ we end up with $
	I_c \sim 10^{-10} - 10^{-6}\text{ }A $,  what is accessible in modern experiments.
If we take the transparency of real junctions into account, we  get the estimates of the same order of magnitude. This generalization is discussed in the next section. It is worth noting that these critical currents are smaller than the ones obtained in a S/F/S structure with the LRTC produced by the Rashba SOC, the exchange field and the condensate motion \cite{silaev2020odd, Bobkova2021} because the typical Rashba constants $\alpha_R \sim 0.1 - 1$ \cite{banerjee2018controlling, lo2014spin, ast2007giant} are larger than $\theta$ and $\kappa$.

In our case the junction is in the $\pi$-state independent of its width $d_F \gg \xi_F$.  The similar $\pi$-shift in the ground state of JJs, where the Josephson coupling is realized via the LRTC, has already been reported in Ref.~\cite{fominov2007josephson}. The estimated amplitude of the critical current is also of the same order  of magnitude as in SFS junctions with LRTC  generated by the magnetic inhomogeneity \cite{fominov2007josephson}.

\begin{figure}[htp]
	\centering
	\includegraphics[width=0.9\linewidth]{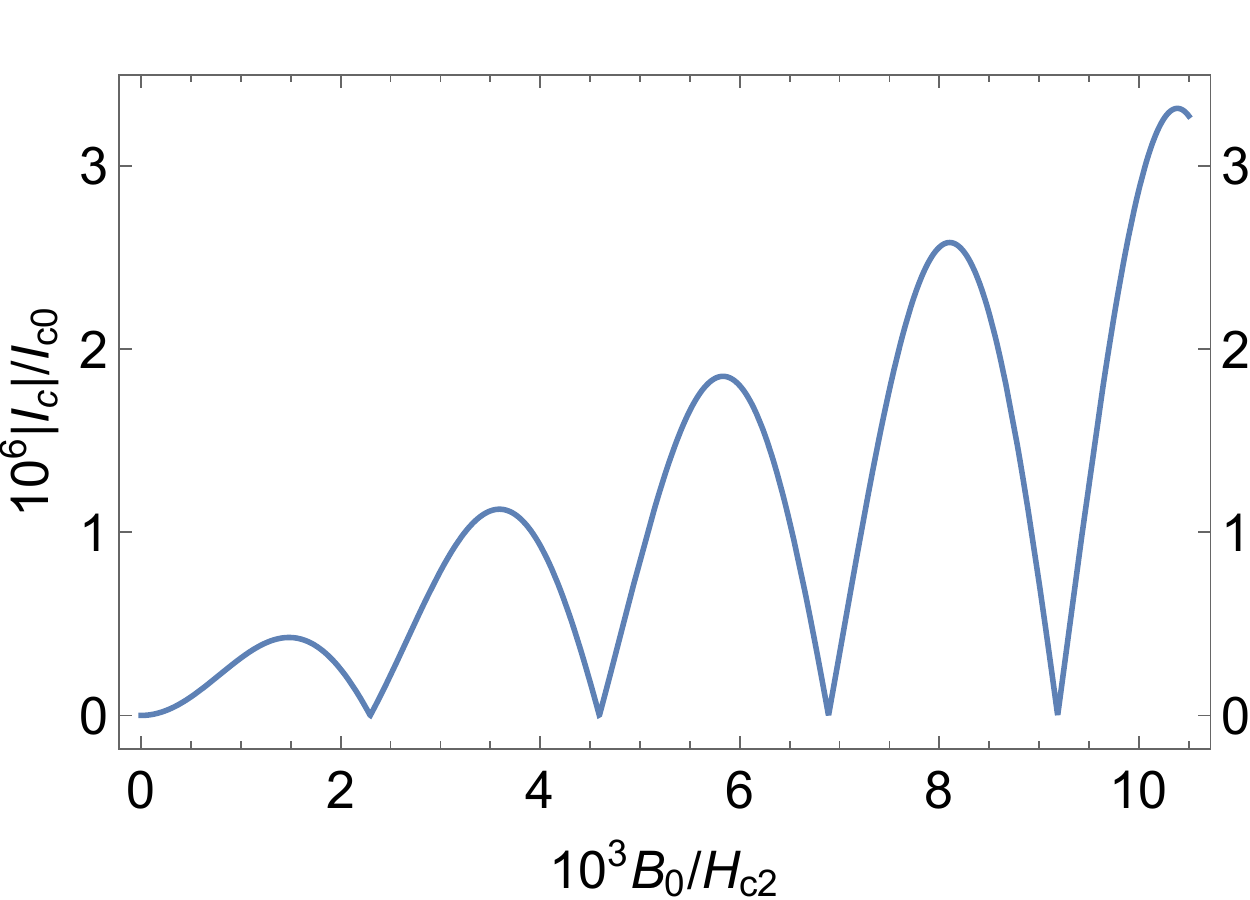}
	\caption{\label{fig:Ic_ideal} Dependence of the critical current  $\vert I_c \vert$ on  $B_0$, when $\mathbf{B} \perp \mathbf{h}$. The parameters are $\theta$ = $\kappa = 0.1$, $h = 50 T_c$, $d_F = 5 \xi_F$, $L = 10 \xi_S$, $\lambda_L = 10 \xi_S$.}
\end{figure}

In order to demonstrate the dependence of the critical current (\ref{eq:JJCDideal}) on the external magnetic field $\mathbf{B}$ we have calculated the superconducting momentum $\mathbf{p}_s$ as a function of $\mathbf{B}$. The calculation is provided in the Appendix.  Under the condition $\lambda_L \gg \xi_S$, we neglect the spatial variation of the condensate momentum in the leads, what imposes $p_{sz} = \frac{2 e \lambda_L}{c}B_{0y} $. We introduce the angle $\alpha$ between the $\hat{\mathbf{e}}_z$ axis and the magnetic field, what means $\mathbf{B} =\hat{\mathbf{e}}_y B_0 \sin{\alpha}  + \hat{\mathbf{e}}_z B_0 \cos{\alpha}$, $p_{sz} = \frac{2 e \lambda_L}{c} B_{0}\sin{\alpha}$ and $p_s = \frac{2 e \lambda_L}{c} B_{0}$. It is seen from Eq.~(\ref{eq:JJCDideal}) that $I_c \sim p^2_{sz}$ reaches its maximal value at $\alpha = \pi/2$. The dependence of the critical current $\frac{\vert I_{c}\vert}{I_{c0}}(\alpha = \pi/2)$ on the absolute value of the external field $B_0$ is shown in Fig.~\ref{fig:Ic_ideal}. The usual phase variation along the junction $\sin{\phi}/\phi$, $\phi = 2\pi \Phi/\Phi_0$, $\Phi = (2 \lambda_L + d_F) L B_0$, $\Phi_0 = \frac{\pi c}{e}$ is taken into account here, what leads to the interference pattern. At small fields the current increases $\propto B^2$.  At larger fields the oscillating critical current has a linear in magnetic field envelope. This trend is bounded from above by the orbital depairing effect and by the vortex penetration into the junction.  The latter happens when the condensate momentum reaches the critical value $p_s \xi_S \sim 1$ at the interfaces \cite{silaev2020odd}, what gives $\frac{B_{c1}}{H_{c2}} \sim \frac{\xi^2_S}{\lambda^2_L} \sim 0.01$ for $\lambda_L = 10\xi_S$. 

\subsection{S/F/S Josephson junction with low-transparent interfaces}
\label{SFS_low}

In this section we consider the opposite case of low-transparent S/F interfaces. The finite transparency of the interfaces is taken into account via the modified Kupriyanov-Lukichev boundary conditions \cite{bergeret2016manifestation}:
\begin{equation}  
	\label{eq:CounOfCurrents}
	\frac{\sigma_S}{D_S} \check{J}_{SL} = \frac{\sigma_F}{D_F}\check{J}_{FL}\text{, }\frac{\sigma_S}{D_S} \check{J}_{SR} = \frac{\sigma_F}{D_F}\check{J}_{FR};\\
\end{equation}
\begin{equation}
\begin{aligned}
	\label{eq:TunnelJ}
	\check{J}_{FL} = \frac{D_F}{2 R_b \sigma_F}\left[\check{g}_{SL}\text{, }\check{g}_{FL}\right]\text{, }
	\\\check{J}_{FR} = -\frac{D_F}{2 R_b \sigma_F}\left[\check{g}_{SR}\text{, }\check{g}_{FR}\right].
\end{aligned}
\end{equation}
$R_b$ is the barrier resistance per unit area of the junction and $\sigma_F$ is the conductivity of the ferromagnetic region. The coefficients in Eq.~(\ref{eq:CounOfCurrents}) result from the continuity of charge and spin currents. We imply $\sigma_S \gg \sigma_F$, where $\sigma_S$ is the conductivity of the superconducting leads. Under this assumption one can neglect the inverse proximity effect in the superconducting leads. The boundary condition Eq.~(\ref{eq:CounOfCurrents}) is reduced to $\check J_{SL,R} = 0$, what corresponds to the approximation of an impenetrable superconducting surface. Substituting the anomalous Green's function in the superconductor (\ref{eq:UsadelSSol}) into the linearized version of this equation, we obtain that the anomalous Green's function in the left superconducting lead up to the first order with respect to $\theta$ and $\kappa$ takes the form:
\begin{equation} \label{eq:STunnelLeft}
\begin{aligned}
	{f}_{SL}=	\begin{pmatrix}
		f_{SLs}\\
		f_{SLx}\\
		f_{SLy}\\
		f_{SLz}
	\end{pmatrix} = \begin{pmatrix}
		-\frac{i\vert \Delta \vert e^{i \varphi/2}}{\vert\omega_n\vert \iffalse + D_S p^2_s/2 \fi}\\
		0\\
		0\\
		0\\
	\end{pmatrix} +\\ + \frac{ \vert \Delta \vert e^{i \varphi/2} }{\vert\omega_n\vert \iffalse + D_S p^2_s/2 \fi}  \frac{ \theta \operatorname{sgn}{\omega_n}}{\lambda_{Ss}}
	\begin{pmatrix}
		0\\
		0\\
		-p_{sz}\\
		p_{sy}\\
	\end{pmatrix} e^{\lambda_{Ss} x}, \text{ for } x\leq 0.
\end{aligned}
\end{equation}
Here it is assumed that $\tau_{so} \gg 1/T_c$, what imposes $\lambda_{St} = \lambda_{Ss}$. For the right lead the calculations are similar. Noting that $p_{sL(z,y)} = - p_{sR(z,y)}$ we obtain
\begin{equation}\label{eq:STunnelRight}
\begin{aligned}
	{f}_{SR}=	\begin{pmatrix}
		f_{SRs}\\
		f_{SRx}\\
		f_{SRy}\\
		f_{SRz}
	\end{pmatrix} = \begin{pmatrix}
		-\frac{i\vert \Delta \vert e^{-i \varphi/2}}{\vert\omega_n\vert \iffalse + D_S p^2_s/2 \fi}\\
		0\\
		0\\
		0\\
	\end{pmatrix}  +\\+ \frac{ \vert \Delta \vert e^{-i \varphi/2}}{\vert\omega_n\vert \iffalse + D_S p^2_s/2 \fi}  \frac{ \theta \operatorname{sgn}{\omega_n}}{\lambda_{Ss}}
	\begin{pmatrix}
		0\\
		0\\
		-p_{sz}\\
		p_{sy}\\
	\end{pmatrix} e^{-\lambda_{Ss} (x - d_F)}, \text{ for } x\geq d_F.
\end{aligned}
\end{equation}
Eqs.~(\ref{eq:STunnelLeft}) and (\ref{eq:STunnelRight}) are instructive. They do not contain $\kappa$. It illustrates once again that $\kappa$ does not produce triplets by its own, at least in the framework of the considered approximation. The triplets are produced by $\theta$. We are interested in the LRTC, which is presented only by $f_{y}$ in the case under consideration, when $\bm h = h \bm e_z$. Making use of Eq.~(\ref{eq:UsadelLinInF}) in the interlayer and linearized version of boundary conditions (\ref{eq:TunnelJ}), where the Green's functions of the superconducting leads are substituted from Eqs.~(\ref{eq:STunnelLeft}) and (\ref{eq:STunnelRight}), we obtain 
\begin{equation}\label{eq:LRTCtunnel}
\begin{aligned}
	f_y = C  \times 
	\Bigl[(f_{SRy} + f_{SLy} e^{\lambda_{FL}d_F}) e^{- \lambda_{FL} x} +  \\  + (f_{SLy} + f_{SRy} e^{\lambda_{FL}d_F}) e^{\lambda_{FL}(x - d_F)}  \Bigr].
\end{aligned}
\end{equation}
Eq.~(\ref{eq:LRTCtunnel}) is obtained up to the first order in the parameter $\gamma  \xi_{FL}$, where $\gamma = [R_b \sigma_F]^{-1}$ accounts for the S/F interface transparency, which is assumed to be low, and $\xi_{FL} = \sqrt{D_F/(2\pi T_c)}$.  $C = \gamma/(2\lambda_{FL}  \sinh{\lambda_{FL} d_F})$. 

Now we are ready to calculate the Josephson current
\begin{equation}\label{eq:JJCDtunnel}
\begin{aligned}
	j_x(\varphi) = i \frac{\sigma_F \pi T_c}{e} \sum_{\omega>0} \left[ \tilde{f}_{y} \partial_x f_{y} - \operatorname{c.c.} \right] =
	\\= - \frac{4 \sigma_F}{e} \frac{\vert\Delta\vert^2}{\pi T_c} \iffalse \frac{1}{\left[1 + \frac{D_S p^2_s}{2 \pi T_c}\right]^2} \fi \frac{1}{2\sinh{\frac{d_F}{\xi_{FL}}}} \xi_{FL}\left({\gamma \theta p_{sz} \xi_S}\right)^2 \sin{\varphi}.
\end{aligned}
\end{equation}
Again, only the lowest Matsubara frequency is taken into account in the bottom line of Eq.~(\ref{eq:JJCDtunnel}). For a short junction $\lambda_{FL} d_F \ll 1$ \iffalse, the small condensate momentum $(p_s \xi_S)^2 \ll 1$, where $\xi_S = \sqrt{\frac{D_S}{2 \pi T_c}}$, \fi Eq.~(\ref{eq:JJCDtunnel}) takes the form:
\begin{align} \label{eq:estimationTunnel}
	j_x(\varphi) &= - \left\{\frac{4 \sigma_F}{e}\frac{\vert\Delta\vert^2}{\pi T_c} \frac{1}{d_F}\right\} \frac{(\gamma \xi_{FL})^2 (\theta p_{sz} \xi_S)^2}{2}  \sin\varphi = \nonumber \\
	&= - j_c \sin\varphi.
\end{align}
Taking data used in the previous section and additionally assuming $\xi_{FL} \approx \xi_S \sim 30\text{ }nm$ and $\gamma \xi_{FL} \sim 0.2$ \cite{silaev2020odd, fominov2007josephson} we find $I_c = j_c S\sim 10^{-10} - 10^{-6}\text{ }A$. We have met an unusual situation, when the critical currents for the junction with ideal interfaces and in the tunnel limit are of the same order of magnitude. The reason can be explained as follows. The main source of the LRTC is the singlet correlations near the interfaces. For the system with transparent interfaces the inverse proximity effect in the superconducting leads is essential. As a consequence the singlet correlations near the interfaces are suppressed, what results in the factor $\xi_F/\xi_S$ in the LRTC (\ref{eq:FSolLinInThetaKappa_xy}) for the system with transparent interfaces. However, in the tunnel limit the singlet correlations near the interfaces practically do not feel the ferromagnetic layer. Therefore, they have the maximal possible value and the intensity of conversion of the singlet correlations into the LRTC is determined only by the transparency resulting in the factor $(\gamma \xi_{FL})$ in (\ref{eq:LRTCtunnel}). So, the mismatch between the ideal case and the tunnel one results from these two factors, which can be of the same order of magnitude in real setups.

\section{Controllable by supercurrent \texorpdfstring{$0-\pi$}{} transitions in the S/F/S Josephson junction}
\label{controllable}

\begin{figure}[htp]
    \centering
    \includegraphics[width=0.9\linewidth]{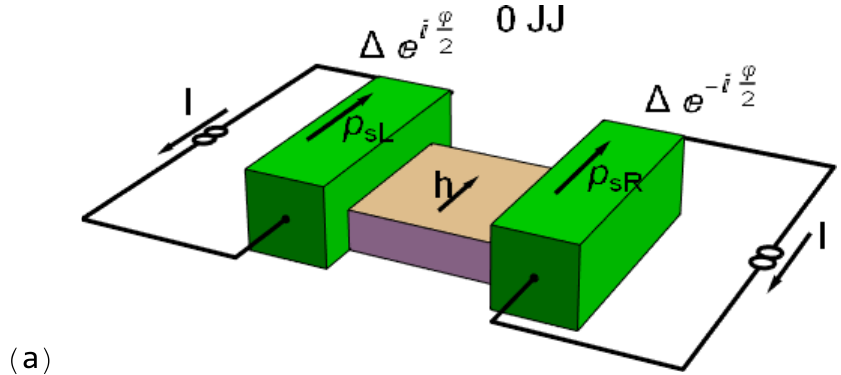}\\
    \includegraphics[width=0.9\linewidth]{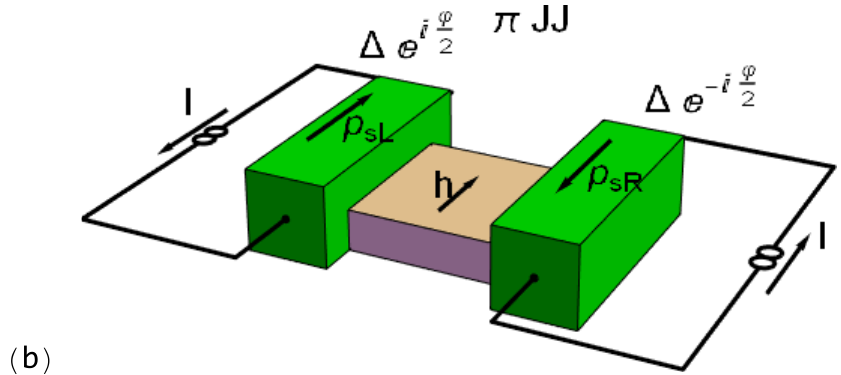}
    \caption{\label{fig:SFS_0_pi} Principle of $0-\pi$ switching due to control of the condensate motion. The condensate motion is produced by the supercurrent, applied in the leads along the S/F interfaces. The direction of the condensate momentum in each of the leads can be reversed by reversing the current.}
\end{figure}

As it has been already mentioned, if the LRTC are generated by the external magnetic field, the corresponding JJ is in the $\pi$-ground state. This fact is based on  the relation $p_{szL} = -p_{szR}$ between condensate momentum directions in the leads. However, the supercurrent along the S/F interfaces is not necessary caused by the magnetic field. It can be applied directly by a current source, see Fig.~\ref{fig:SFS_0_pi}. In this case the absolute value and the direction of $\bm p_s$ in each of the leads is determined by the applied current. In particular, it is possible to switch between the situations $p_{szL} = p_{szR}$ and $p_{szL} = - p_{szR}$ by reversing the direction of the applied current in one of the leads. The switching results in the $0-\pi$ transition of the ground state phase of the JJ. Indeed, it can be shown that the  generalization of Eq.~(\ref{eq:JJCDideal}) to the case of arbitrary $\bm p_{sL(R)}$, which do not obey  condition $p_{szL} = -p_{szR}$, takes the form: \begin{equation} \label{eq:JJCDideal1}
\begin{aligned}
	j_x(\varphi) = \frac{4 \sigma }{e} \frac{\vert \Delta\vert^2}{\pi T_c} \iffalse{\left[\pi T_c + \frac{D p^2_s}{2}\right]^2} \fi \frac{\left\{\kappa^2 + \theta^2\left(1 + \frac{\xi_F}{\xi_S}\right)^2 \right\}}{\left[1 + \left(1 + \frac{\xi_F}{\xi_S}\right)^2\right]^2}  \frac{\xi^2_F}{4\xi_S} e^{-d_F/\xi_S} \times \\ \times \left[ (\mathbf{n}_{\mathbf{h}}\cdot \mathbf{p}_{sR}) ((\mathbf{n}_{\mathbf{h}}\cdot \mathbf{p}_{sL})\right] \sin{\varphi}.
\end{aligned}
\end{equation}
Here $\mathbf{n}_{\mathbf{h}} = \mathbf{h}/h$. While for the case of low-transparent interfaces the corresponding generalization of Eq.~(\ref{eq:JJCDtunnel}) reads as follows:
\begin{equation}
\begin{aligned} \label{eq:JJCDtunnel1}
	j_x(\varphi) = \frac{4 \sigma_F}{e} \frac{\vert\Delta\vert^2}{\pi T_c} \iffalse \frac{1}{\left[1 + \frac{D_S p^2_s}{2 \pi T_c}\right]^2} \fi \frac{1}{2\sinh{\frac{d_F}{\xi_{FL}}}} \xi_{FL}\left({ \gamma \theta \xi_S}\right)^2 \times \\ \times \left[(\mathbf{n}_{\mathbf{h}}\cdot \mathbf{p}_{sR}) (\mathbf{n}_{\mathbf{h}}\cdot \mathbf{p}_{sL}) \right] \sin{\varphi}.
\end{aligned}    
\end{equation}

It is seen that the sign of the critical current is controlled by the product  $\left[  (\mathbf{n}_{\mathbf{h}}\cdot \mathbf{p}_{sR}) (\mathbf{n}_{\mathbf{h}}\cdot \mathbf{p}_{sL}) \right]$. Therefore, it is possible to switch the ground state of the JJ between $0$ and $\pi$ by reversing the direction of the supercurrent in the leads. In the proposed geometry the width of the S/F interface along the supercurrent direction assumed to be small enough in order to neglect the superconducting phase variation along the interface.

It is interesting that the effect of the controllable $0-\pi$ phase transition is rather general and can be realized not only for JJs with extrinsic SOC in the leads, but also for JJs with Rashba-type SOC at the interfaces. The interface Rashba SOC can be both of intrinsic nature or due to the additional, for example, Pt layers at the S/F interfaces. For interfaces with Rashba SOC the  effective boundary condition was introduced in Ref.~\cite{silaev2020odd}
\begin{equation} \label{eq:effective_bound_cond}
\begin{cases} 
	(\mathbf{n}\cdot\nabla) \check{f}_s = \gamma \check{F}_{BCS},\\
	(\mathbf{n}\cdot\nabla) \check{\mathbf{f}} = 4i (\alpha d_{so}) \check{\tau}_3\check{\mathbf{f}}\times\left[\mathbf{p}_s\times \mathbf{n}\right].
\end{cases}
\end{equation}
Here $\check f_s = f_s \check \tau_+ - f_s^c \check \tau_-$ with $\check \tau_\pm = (\check \tau_1 \pm i \check \tau_2)/2 $ and $\check{\mathbf{f}} =  {\mathbf{f}}_t \check \tau_+ -  {\mathbf{f}}_t^c \check \tau_-$. $\check{F}_{BCS} = -i  \check{\Delta}/\vert \omega_n\vert$, $\alpha$ is the Rashba SOC constant, $d_{so}$ is a width of the Rashba SOC layer, $\mathbf{n} = (\pm1\text{, }0\text{, }0)$ is a vector perpendicular to the interface.

The Green's function in the ferromagnetic layer is determined by the Usadel equation (\ref{eq:UsadelLinInF}) supplemented by the boundary conditions (\ref{eq:effective_bound_cond}). Further calculations are performed up to the first order with respect to $\tilde{\alpha} = \alpha d_{so} \ll 1$. Up to the zero order with respect to this parameter there are only short-range triplet correlations in the system and at the S/F interfaces the corresponding anomalous Green's functions take the form:
\begin{equation}
    f_{z}(0) =  \frac{\gamma \xi_F}{2}  \frac{\vert\Delta\vert e^{i \varphi/2}}{\omega_n} \text{, } f_{z}(d_F) =  \frac{\gamma \xi_F}{2} \frac{\vert\Delta\vert e^{-i \varphi/2}}{\omega_n},
\end{equation}
where we have taken into account that $d_F \gg \xi_F$. 

The LRTC are generated by the short-range triplets $f_{z}$ and the only nonzero component of the LTRC is $f_{x}$, which takes the form:
\begin{equation}
\begin{aligned}
    f_{x}(x) =- \frac{2 i \tilde{\alpha} \gamma \xi_F }{\lambda_{FL}\sinh \lambda_{FL} d_F }  \frac{\vert\Delta\vert}{ \omega_n }\times \\ \times \left\{ -p_{szL}e^{i\varphi/2}\cosh{\lambda_{FL} (d_F - x)} + \right. \\ \left.  + p_{szR} e^{-i\varphi/2} \cosh{\lambda_{FL} x} \right\}.
\end{aligned}
\end{equation}

Then the Josephson current through the system reads
\begin{equation}
\begin{aligned}
	j_x(\varphi) =  i \frac{\sigma_F \pi T_c}{e} \sum_{\omega>0} \left[\tilde{f_x} \partial_x f_x - c.c.\right] =
	\\ = \frac{4 \sigma_F}{e} \frac{\vert\Delta\vert^2}{\pi T_c} \frac{2  (\tilde{\alpha} \gamma \xi_F)^2
 \xi_{FL} }{\sinh{\frac{d_F}{\xi_{FL}}}} \left[ (\mathbf{n}_{\mathbf{h}}\cdot \mathbf{p}_{sR}) (\mathbf{n}_{\mathbf{h}}\cdot \mathbf{p}_{sL}) \right] \sin{\varphi}.
\end{aligned}
\end{equation}
Here we have incorporated the fact that $p_{sz(L,R)} = (\mathbf{n}_{\mathbf{h}}\cdot \mathbf{p}_{s(L,R)})$. It is seen that the current is again proportional to the product  $\left[ (\mathbf{n}_{\mathbf{h}}\cdot \mathbf{p}_{sR}) (\mathbf{n}_{\mathbf{h}}\cdot \mathbf{p}_{sL}) \right]$, what demonstrates that the system is also suitable for the controllable $0-\pi$ switching.

\section{Conclusions}
\label{conclusions}

It is predicted that the impurity SOC together with the condensate motion along the superconducting leads induces long-range spin-triplet superconductivity in superconductor/ferromagnet heterostructures. This effect has two microscopic mechanisms. The first one is the generation of the triplet pairs in the superconductor via the superconducting  spin Hall effect. Under the appropriate choice of the condensate momentum direction in plane of the S/F interfaces the pairs are long-range in the ferromagnet. The second mechanism is associated with the singlet-triplet conversion at the S/F interface with subsequent rotation of spin of the resulting short-range triplet pairs via the spin current swapping effect. The corresponding long-range triplet anomalous Green's functions in S/F and S/F/S hybrids have been calculated. They result in the long-range Josephson effect in S/F/S JJs and the corresponding  current-phase relations have been obtained for JJs  with both high- and low-transparent interfaces. Our finding expands the range of systems, where the effect of the LRTC generation by the supercurrent is possible. We believe that the effect opens new avenue in superconducting electronics because allows for a total low-dissipative control of the amplitude and ground state phase of the S/F/S JJs. Namely, the Josephson critical current can be switched on/off by the supercurrent motion along the interfaces and ground state phase of the junction can be switched between $0$ and $\pi$ states by reversing the direction of the supercurrent in one of the leads.

\begin{acknowledgments}
We are thankful to Mikhail Silaev for valuable discussions. The work of I.V.B has been carried out within the state task of ISSP RAS with the support by RFBR grant 19-02-00466. I.V.B. and A.A.M. also acknowledge the financial support by Foundation for the Advancement of Theoretical Physics and Mathematics “BASIS”.
\end{acknowledgments}
 
\bibliographystyle{apsrev4-2}
\bibliography{refs_SOtriplets}

\begin{thebibliography}{87}%
\makeatletter
\providecommand \@ifxundefined [1]{%
 \@ifx{#1\undefined}
}%
\providecommand \@ifnum [1]{%
 \ifnum #1\expandafter \@firstoftwo
 \else \expandafter \@secondoftwo
 \fi
}%
\providecommand \@ifx [1]{%
 \ifx #1\expandafter \@firstoftwo
 \else \expandafter \@secondoftwo
 \fi
}%
\providecommand \natexlab [1]{#1}%
\providecommand \enquote  [1]{``#1''}%
\providecommand \bibnamefont  [1]{#1}%
\providecommand \bibfnamefont [1]{#1}%
\providecommand \citenamefont [1]{#1}%
\providecommand \href@noop [0]{\@secondoftwo}%
\providecommand \href [0]{\begingroup \@sanitize@url \@href}%
\providecommand \@href[1]{\@@startlink{#1}\@@href}%
\providecommand \@@href[1]{\endgroup#1\@@endlink}%
\providecommand \@sanitize@url [0]{\catcode `\\12\catcode `\$12\catcode
  `\&12\catcode `\#12\catcode `\^12\catcode `\_12\catcode `\%12\relax}%
\providecommand \@@startlink[1]{}%
\providecommand \@@endlink[0]{}%
\providecommand \url  [0]{\begingroup\@sanitize@url \@url }%
\providecommand \@url [1]{\endgroup\@href {#1}{\urlprefix }}%
\providecommand \urlprefix  [0]{URL }%
\providecommand \Eprint [0]{\href }%
\providecommand \doibase [0]{https://doi.org/}%
\providecommand \selectlanguage [0]{\@gobble}%
\providecommand \bibinfo  [0]{\@secondoftwo}%
\providecommand \bibfield  [0]{\@secondoftwo}%
\providecommand \translation [1]{[#1]}%
\providecommand \BibitemOpen [0]{}%
\providecommand \bibitemStop [0]{}%
\providecommand \bibitemNoStop [0]{.\EOS\space}%
\providecommand \EOS [0]{\spacefactor3000\relax}%
\providecommand \BibitemShut  [1]{\csname bibitem#1\endcsname}%
\let\auto@bib@innerbib\@empty
%</preamble>
\bibitem [{\citenamefont {Clark}\ \emph {et~al.}(1980)\citenamefont {Clark},
  \citenamefont {Prance},\ and\ \citenamefont {Grassie}}]{Clark1980}%
  \BibitemOpen
  \bibfield  {author} {\bibinfo {author} {\bibfnamefont {T.~D.}\ \bibnamefont
  {Clark}}, \bibinfo {author} {\bibfnamefont {R.~J.}\ \bibnamefont {Prance}},\
  and\ \bibinfo {author} {\bibfnamefont {A.~D.~C.}\ \bibnamefont {Grassie}},\
  }\href {https://doi.org/10.1063/1.327935} {\bibfield  {journal} {\bibinfo
  {journal} {Journal of Applied Physics}\ }\textbf {\bibinfo {volume} {51}},\
  \bibinfo {pages} {2736} (\bibinfo {year} {1980})}\BibitemShut {NoStop}%
\bibitem [{\citenamefont {Yamashita}\ \emph {et~al.}(2005)\citenamefont
  {Yamashita}, \citenamefont {Tanikawa}, \citenamefont {Takahashi},\ and\
  \citenamefont {Maekawa}}]{Yamashita2005}%
  \BibitemOpen
  \bibfield  {author} {\bibinfo {author} {\bibfnamefont {T.}~\bibnamefont
  {Yamashita}}, \bibinfo {author} {\bibfnamefont {K.}~\bibnamefont {Tanikawa}},
  \bibinfo {author} {\bibfnamefont {S.}~\bibnamefont {Takahashi}},\ and\
  \bibinfo {author} {\bibfnamefont {S.}~\bibnamefont {Maekawa}},\ }\href
  {https://doi.org/10.1103/PhysRevLett.95.097001} {\bibfield  {journal}
  {\bibinfo  {journal} {Phys. Rev. Lett.}\ }\textbf {\bibinfo {volume} {95}},\
  \bibinfo {pages} {097001} (\bibinfo {year} {2005})}\BibitemShut {NoStop}%
\bibitem [{\citenamefont {Feofanov}\ \emph {et~al.}(2010)\citenamefont
  {Feofanov}, \citenamefont {Oboznov}, \citenamefont {Bol'ginov}, \citenamefont
  {Lisenfeld}, \citenamefont {Poletto}, \citenamefont {Ryazanov}, \citenamefont
  {Rossolenko}, \citenamefont {Khabipov}, \citenamefont {Balashov},
  \citenamefont {Zorin}, \citenamefont {Dmitriev}, \citenamefont {Koshelets},\
  and\ \citenamefont {Ustinov}}]{Feofanov2010}%
  \BibitemOpen
  \bibfield  {author} {\bibinfo {author} {\bibfnamefont {A.~K.}\ \bibnamefont
  {Feofanov}}, \bibinfo {author} {\bibfnamefont {V.~A.}\ \bibnamefont
  {Oboznov}}, \bibinfo {author} {\bibfnamefont {V.~V.}\ \bibnamefont
  {Bol'ginov}}, \bibinfo {author} {\bibfnamefont {J.}~\bibnamefont
  {Lisenfeld}}, \bibinfo {author} {\bibfnamefont {S.}~\bibnamefont {Poletto}},
  \bibinfo {author} {\bibfnamefont {V.~V.}\ \bibnamefont {Ryazanov}}, \bibinfo
  {author} {\bibfnamefont {A.~N.}\ \bibnamefont {Rossolenko}}, \bibinfo
  {author} {\bibfnamefont {M.}~\bibnamefont {Khabipov}}, \bibinfo {author}
  {\bibfnamefont {D.}~\bibnamefont {Balashov}}, \bibinfo {author}
  {\bibfnamefont {A.~B.}\ \bibnamefont {Zorin}}, \bibinfo {author}
  {\bibfnamefont {P.~N.}\ \bibnamefont {Dmitriev}}, \bibinfo {author}
  {\bibfnamefont {V.~P.}\ \bibnamefont {Koshelets}},\ and\ \bibinfo {author}
  {\bibfnamefont {A.~V.}\ \bibnamefont {Ustinov}},\ }\href
  {https://doi.org/10.1038/nphys1700} {\bibfield  {journal} {\bibinfo
  {journal} {Nature Physics}\ }\textbf {\bibinfo {volume} {6}},\ \bibinfo
  {pages} {593} (\bibinfo {year} {2010})}\BibitemShut {NoStop}%
\bibitem [{\citenamefont {Shcherbakova}\ \emph {et~al.}(2015)\citenamefont
  {Shcherbakova}, \citenamefont {Fedorov}, \citenamefont {Shulga},
  \citenamefont {Ryazanov}, \citenamefont {Bolginov}, \citenamefont {Oboznov},
  \citenamefont {Egorov}, \citenamefont {Shkolnikov}, \citenamefont {Wolf},
  \citenamefont {Beckmann},\ and\ \citenamefont {Ustinov}}]{Shcherbakova2015}%
  \BibitemOpen
  \bibfield  {author} {\bibinfo {author} {\bibfnamefont {A.~V.}\ \bibnamefont
  {Shcherbakova}}, \bibinfo {author} {\bibfnamefont {K.~G.}\ \bibnamefont
  {Fedorov}}, \bibinfo {author} {\bibfnamefont {K.~V.}\ \bibnamefont {Shulga}},
  \bibinfo {author} {\bibfnamefont {V.~V.}\ \bibnamefont {Ryazanov}}, \bibinfo
  {author} {\bibfnamefont {V.~V.}\ \bibnamefont {Bolginov}}, \bibinfo {author}
  {\bibfnamefont {V.~A.}\ \bibnamefont {Oboznov}}, \bibinfo {author}
  {\bibfnamefont {S.~V.}\ \bibnamefont {Egorov}}, \bibinfo {author}
  {\bibfnamefont {V.~O.}\ \bibnamefont {Shkolnikov}}, \bibinfo {author}
  {\bibfnamefont {M.~J.}\ \bibnamefont {Wolf}}, \bibinfo {author}
  {\bibfnamefont {D.}~\bibnamefont {Beckmann}},\ and\ \bibinfo {author}
  {\bibfnamefont {A.~V.}\ \bibnamefont {Ustinov}},\ }\href
  {https://doi.org/10.1088/0953-2048/28/2/025009} {\bibfield  {journal}
  {\bibinfo  {journal} {Superconductor Science and Technology}\ }\textbf
  {\bibinfo {volume} {28}},\ \bibinfo {pages} {025009} (\bibinfo {year}
  {2015})}\BibitemShut {NoStop}%
\bibitem [{\citenamefont {Buzdin}\ \emph {et~al.}(1982)\citenamefont {Buzdin},
  \citenamefont {Bulaevskii},\ and\ \citenamefont {V.}}]{Buzdin1982}%
  \BibitemOpen
  \bibfield  {author} {\bibinfo {author} {\bibfnamefont {A.~I.}\ \bibnamefont
  {Buzdin}}, \bibinfo {author} {\bibfnamefont {L.~N.}\ \bibnamefont
  {Bulaevskii}},\ and\ \bibinfo {author} {\bibfnamefont {P.~S.}\ \bibnamefont
  {V.}},\ }\href {http://jetpletters.ru/ps/1314/article_19853.shtml} {\bibfield
   {journal} {\bibinfo  {journal} {JETP Lett.}\ }\textbf {\bibinfo {volume}
  {35}},\ \bibinfo {pages} {178} (\bibinfo {year} {1982})}\BibitemShut
  {NoStop}%
\bibitem [{\citenamefont {Buzdin}(2005)}]{Buzdin2005}%
  \BibitemOpen
  \bibfield  {author} {\bibinfo {author} {\bibfnamefont {A.~I.}\ \bibnamefont
  {Buzdin}},\ }\href {https://doi.org/10.1103/RevModPhys.77.935} {\bibfield
  {journal} {\bibinfo  {journal} {Rev. Mod. Phys.}\ }\textbf {\bibinfo {volume}
  {77}},\ \bibinfo {pages} {935} (\bibinfo {year} {2005})}\BibitemShut
  {NoStop}%
\bibitem [{\citenamefont {Kontos}\ \emph {et~al.}(2002)\citenamefont {Kontos},
  \citenamefont {Aprili}, \citenamefont {Lesueur}, \citenamefont {Gen\^et},
  \citenamefont {Stephanidis},\ and\ \citenamefont {Boursier}}]{Kontos2002}%
  \BibitemOpen
  \bibfield  {author} {\bibinfo {author} {\bibfnamefont {T.}~\bibnamefont
  {Kontos}}, \bibinfo {author} {\bibfnamefont {M.}~\bibnamefont {Aprili}},
  \bibinfo {author} {\bibfnamefont {J.}~\bibnamefont {Lesueur}}, \bibinfo
  {author} {\bibfnamefont {F.}~\bibnamefont {Gen\^et}}, \bibinfo {author}
  {\bibfnamefont {B.}~\bibnamefont {Stephanidis}},\ and\ \bibinfo {author}
  {\bibfnamefont {R.}~\bibnamefont {Boursier}},\ }\href
  {https://doi.org/10.1103/PhysRevLett.89.137007} {\bibfield  {journal}
  {\bibinfo  {journal} {Phys. Rev. Lett.}\ }\textbf {\bibinfo {volume} {89}},\
  \bibinfo {pages} {137007} (\bibinfo {year} {2002})}\BibitemShut {NoStop}%
\bibitem [{\citenamefont {Ryazanov}\ \emph {et~al.}(2001)\citenamefont
  {Ryazanov}, \citenamefont {Oboznov}, \citenamefont {Rusanov}, \citenamefont
  {Veretennikov}, \citenamefont {Golubov},\ and\ \citenamefont
  {Aarts}}]{Ryazanov2001}%
  \BibitemOpen
  \bibfield  {author} {\bibinfo {author} {\bibfnamefont {V.~V.}\ \bibnamefont
  {Ryazanov}}, \bibinfo {author} {\bibfnamefont {V.~A.}\ \bibnamefont
  {Oboznov}}, \bibinfo {author} {\bibfnamefont {A.~Y.}\ \bibnamefont
  {Rusanov}}, \bibinfo {author} {\bibfnamefont {A.~V.}\ \bibnamefont
  {Veretennikov}}, \bibinfo {author} {\bibfnamefont {A.~A.}\ \bibnamefont
  {Golubov}},\ and\ \bibinfo {author} {\bibfnamefont {J.}~\bibnamefont
  {Aarts}},\ }\href {https://doi.org/10.1103/PhysRevLett.86.2427} {\bibfield
  {journal} {\bibinfo  {journal} {Phys. Rev. Lett.}\ }\textbf {\bibinfo
  {volume} {86}},\ \bibinfo {pages} {2427} (\bibinfo {year}
  {2001})}\BibitemShut {NoStop}%
\bibitem [{\citenamefont {Oboznov}\ \emph {et~al.}(2006)\citenamefont
  {Oboznov}, \citenamefont {Bol'ginov}, \citenamefont {Feofanov}, \citenamefont
  {Ryazanov},\ and\ \citenamefont {Buzdin}}]{Oboznov2006}%
  \BibitemOpen
  \bibfield  {author} {\bibinfo {author} {\bibfnamefont {V.~A.}\ \bibnamefont
  {Oboznov}}, \bibinfo {author} {\bibfnamefont {V.~V.}\ \bibnamefont
  {Bol'ginov}}, \bibinfo {author} {\bibfnamefont {A.~K.}\ \bibnamefont
  {Feofanov}}, \bibinfo {author} {\bibfnamefont {V.~V.}\ \bibnamefont
  {Ryazanov}},\ and\ \bibinfo {author} {\bibfnamefont {A.~I.}\ \bibnamefont
  {Buzdin}},\ }\href {https://doi.org/10.1103/PhysRevLett.96.197003} {\bibfield
   {journal} {\bibinfo  {journal} {Phys. Rev. Lett.}\ }\textbf {\bibinfo
  {volume} {96}},\ \bibinfo {pages} {197003} (\bibinfo {year}
  {2006})}\BibitemShut {NoStop}%
\bibitem [{\citenamefont {Bannykh}\ \emph {et~al.}(2009)\citenamefont
  {Bannykh}, \citenamefont {Pfeiffer}, \citenamefont {Stolyarov}, \citenamefont
  {Batov}, \citenamefont {Ryazanov},\ and\ \citenamefont
  {Weides}}]{bannykh2009josephson}%
  \BibitemOpen
  \bibfield  {author} {\bibinfo {author} {\bibfnamefont {A.~A.}\ \bibnamefont
  {Bannykh}}, \bibinfo {author} {\bibfnamefont {J.}~\bibnamefont {Pfeiffer}},
  \bibinfo {author} {\bibfnamefont {V.~S.}\ \bibnamefont {Stolyarov}}, \bibinfo
  {author} {\bibfnamefont {I.~E.}\ \bibnamefont {Batov}}, \bibinfo {author}
  {\bibfnamefont {V.~V.}\ \bibnamefont {Ryazanov}},\ and\ \bibinfo {author}
  {\bibfnamefont {M.}~\bibnamefont {Weides}},\ }\href
  {https://journals.aps.org/prb/abstract/10.1103/PhysRevB.79.054501} {\bibfield
   {journal} {\bibinfo  {journal} {Physical Review B}\ }\textbf {\bibinfo
  {volume} {79}},\ \bibinfo {pages} {054501} (\bibinfo {year}
  {2009})}\BibitemShut {NoStop}%
\bibitem [{\citenamefont {Robinson}\ \emph {et~al.}(2006)\citenamefont
  {Robinson}, \citenamefont {Piano}, \citenamefont {Burnell}, \citenamefont
  {Bell},\ and\ \citenamefont {Blamire}}]{robinson2006critical}%
  \BibitemOpen
  \bibfield  {author} {\bibinfo {author} {\bibfnamefont {J.~W.~A.}\
  \bibnamefont {Robinson}}, \bibinfo {author} {\bibfnamefont {S.}~\bibnamefont
  {Piano}}, \bibinfo {author} {\bibfnamefont {G.}~\bibnamefont {Burnell}},
  \bibinfo {author} {\bibfnamefont {C.}~\bibnamefont {Bell}},\ and\ \bibinfo
  {author} {\bibfnamefont {M.~G.}\ \bibnamefont {Blamire}},\ }\href
  {https://journals.aps.org/prl/abstract/10.1103/PhysRevLett.97.177003}
  {\bibfield  {journal} {\bibinfo  {journal} {Physical review letters}\
  }\textbf {\bibinfo {volume} {97}},\ \bibinfo {pages} {177003} (\bibinfo
  {year} {2006})}\BibitemShut {NoStop}%
\bibitem [{\citenamefont {Schulz}\ \emph {et~al.}(2000)\citenamefont {Schulz},
  \citenamefont {Chesca}, \citenamefont {G{\"o}tz}, \citenamefont {Schneider},
  \citenamefont {Schmehl}, \citenamefont {Bielefeldt}, \citenamefont
  {Hilgenkamp}, \citenamefont {Mannhart},\ and\ \citenamefont
  {Tsuei}}]{schulz2000design}%
  \BibitemOpen
  \bibfield  {author} {\bibinfo {author} {\bibfnamefont {R.~R.}\ \bibnamefont
  {Schulz}}, \bibinfo {author} {\bibfnamefont {B.}~\bibnamefont {Chesca}},
  \bibinfo {author} {\bibfnamefont {B.}~\bibnamefont {G{\"o}tz}}, \bibinfo
  {author} {\bibfnamefont {C.~W.}\ \bibnamefont {Schneider}}, \bibinfo {author}
  {\bibfnamefont {A.}~\bibnamefont {Schmehl}}, \bibinfo {author} {\bibfnamefont
  {H.}~\bibnamefont {Bielefeldt}}, \bibinfo {author} {\bibfnamefont
  {H.}~\bibnamefont {Hilgenkamp}}, \bibinfo {author} {\bibfnamefont
  {J.}~\bibnamefont {Mannhart}},\ and\ \bibinfo {author} {\bibfnamefont
  {C.}~\bibnamefont {Tsuei}},\ }\href
  {https://aip.scitation.org/doi/abs/10.1063/1.125627} {\bibfield  {journal}
  {\bibinfo  {journal} {Applied Physics Letters}\ }\textbf {\bibinfo {volume}
  {76}},\ \bibinfo {pages} {912} (\bibinfo {year} {2000})}\BibitemShut
  {NoStop}%
\bibitem [{\citenamefont {Smilde}\ \emph {et~al.}(2002)\citenamefont {Smilde},
  \citenamefont {Ariando}, \citenamefont {Blank}, \citenamefont {Gerritsma},
  \citenamefont {Hilgenkamp},\ and\ \citenamefont {Rogalla}}]{smilde2002d}%
  \BibitemOpen
  \bibfield  {author} {\bibinfo {author} {\bibfnamefont {H.~J.~H.}\
  \bibnamefont {Smilde}}, \bibinfo {author} {\bibnamefont {Ariando}}, \bibinfo
  {author} {\bibfnamefont {D.~H.~A.}\ \bibnamefont {Blank}}, \bibinfo {author}
  {\bibfnamefont {G.~J.}\ \bibnamefont {Gerritsma}}, \bibinfo {author}
  {\bibfnamefont {H.}~\bibnamefont {Hilgenkamp}},\ and\ \bibinfo {author}
  {\bibfnamefont {H.}~\bibnamefont {Rogalla}},\ }\href
  {https://journals.aps.org/prl/abstract/10.1103/PhysRevLett.88.057004}
  {\bibfield  {journal} {\bibinfo  {journal} {Physical review letters}\
  }\textbf {\bibinfo {volume} {88}},\ \bibinfo {pages} {057004} (\bibinfo
  {year} {2002})}\BibitemShut {NoStop}%
\bibitem [{\citenamefont {Hilgenkamp}\ \emph {et~al.}(2003)\citenamefont
  {Hilgenkamp}, \citenamefont {Smilde}, \citenamefont {Blank}, \citenamefont
  {Rijnders}, \citenamefont {Rogalla}, \citenamefont {Kirtley}, \citenamefont
  {Tsuei} \emph {et~al.}}]{hilgenkamp2003ordering}%
  \BibitemOpen
  \bibfield  {author} {\bibinfo {author} {\bibfnamefont {H.}~\bibnamefont
  {Hilgenkamp}}, \bibinfo {author} {\bibfnamefont {H.-J.~H.}\ \bibnamefont
  {Smilde}}, \bibinfo {author} {\bibfnamefont {D.~H.}\ \bibnamefont {Blank}},
  \bibinfo {author} {\bibfnamefont {G.}~\bibnamefont {Rijnders}}, \bibinfo
  {author} {\bibfnamefont {H.}~\bibnamefont {Rogalla}}, \bibinfo {author}
  {\bibfnamefont {J.~R.}\ \bibnamefont {Kirtley}}, \bibinfo {author}
  {\bibfnamefont {C.~C.}\ \bibnamefont {Tsuei}}, \emph {et~al.},\ }\href
  {https://www.nature.com/articles/nature01442} {\bibfield  {journal} {\bibinfo
   {journal} {Nature}\ }\textbf {\bibinfo {volume} {422}},\ \bibinfo {pages}
  {50} (\bibinfo {year} {2003})}\BibitemShut {NoStop}%
\bibitem [{\citenamefont {Ariando}\ \emph {et~al.}(2005)\citenamefont
  {Ariando}, \citenamefont {Darminto}, \citenamefont {Smilde}, \citenamefont
  {Leca}, \citenamefont {Blank}, \citenamefont {Rogalla},\ and\ \citenamefont
  {Hilgenkamp}}]{ariando2005phase}%
  \BibitemOpen
  \bibfield  {author} {\bibinfo {author} {\bibnamefont {Ariando}}, \bibinfo
  {author} {\bibfnamefont {D.}~\bibnamefont {Darminto}}, \bibinfo {author}
  {\bibfnamefont {H.~J.~H.}\ \bibnamefont {Smilde}}, \bibinfo {author}
  {\bibfnamefont {V.}~\bibnamefont {Leca}}, \bibinfo {author} {\bibfnamefont
  {D.~H.~A.}\ \bibnamefont {Blank}}, \bibinfo {author} {\bibfnamefont
  {H.}~\bibnamefont {Rogalla}},\ and\ \bibinfo {author} {\bibfnamefont
  {H.}~\bibnamefont {Hilgenkamp}},\ }\href
  {https://journals.aps.org/prl/abstract/10.1103/PhysRevLett.94.167001}
  {\bibfield  {journal} {\bibinfo  {journal} {Physical review letters}\
  }\textbf {\bibinfo {volume} {94}},\ \bibinfo {pages} {167001} (\bibinfo
  {year} {2005})}\BibitemShut {NoStop}%
\bibitem [{\citenamefont {Lombardi}\ \emph {et~al.}(2002)\citenamefont
  {Lombardi}, \citenamefont {Tafuri}, \citenamefont {Ricci}, \citenamefont
  {Miletto~Granozio}, \citenamefont {Barone}, \citenamefont {Testa},
  \citenamefont {Sarnelli}, \citenamefont {Kirtley},\ and\ \citenamefont
  {Tsuei}}]{lombardi2002intrinsic}%
  \BibitemOpen
  \bibfield  {author} {\bibinfo {author} {\bibfnamefont {F.}~\bibnamefont
  {Lombardi}}, \bibinfo {author} {\bibfnamefont {F.}~\bibnamefont {Tafuri}},
  \bibinfo {author} {\bibfnamefont {F.}~\bibnamefont {Ricci}}, \bibinfo
  {author} {\bibfnamefont {F.}~\bibnamefont {Miletto~Granozio}}, \bibinfo
  {author} {\bibfnamefont {A.}~\bibnamefont {Barone}}, \bibinfo {author}
  {\bibfnamefont {G.}~\bibnamefont {Testa}}, \bibinfo {author} {\bibfnamefont
  {E.}~\bibnamefont {Sarnelli}}, \bibinfo {author} {\bibfnamefont {J.~R.}\
  \bibnamefont {Kirtley}},\ and\ \bibinfo {author} {\bibfnamefont {C.~C.}\
  \bibnamefont {Tsuei}},\ }\href
  {https://journals.aps.org/prl/pdf/10.1103/PhysRevLett.89.207001} {\bibfield
  {journal} {\bibinfo  {journal} {Physical review letters}\ }\textbf {\bibinfo
  {volume} {89}},\ \bibinfo {pages} {207001} (\bibinfo {year}
  {2002})}\BibitemShut {NoStop}%
\bibitem [{\citenamefont {Baselmans}\ \emph {et~al.}(1999)\citenamefont
  {Baselmans}, \citenamefont {Morpurgo}, \citenamefont {van Wees},\ and\
  \citenamefont {Klapwijk}}]{Baselmans1999}%
  \BibitemOpen
  \bibfield  {author} {\bibinfo {author} {\bibfnamefont {J.~J.~A.}\
  \bibnamefont {Baselmans}}, \bibinfo {author} {\bibfnamefont {A.~F.}\
  \bibnamefont {Morpurgo}}, \bibinfo {author} {\bibfnamefont {B.~J.}\
  \bibnamefont {van Wees}},\ and\ \bibinfo {author} {\bibfnamefont {T.~M.}\
  \bibnamefont {Klapwijk}},\ }\href {https://doi.org/10.1038/16204} {\bibfield
  {journal} {\bibinfo  {journal} {Nature}\ }\textbf {\bibinfo {volume} {397}},\
  \bibinfo {pages} {43} (\bibinfo {year} {1999})}\BibitemShut {NoStop}%
\bibitem [{\citenamefont {Golikova}\ \emph {et~al.}(2021)\citenamefont
  {Golikova}, \citenamefont {Wolf}, \citenamefont {Beckmann}, \citenamefont
  {Penzyakov}, \citenamefont {Batov}, \citenamefont {Bobkova}, \citenamefont
  {Bobkov},\ and\ \citenamefont {Ryazanov}}]{golikova2021controllable}%
  \BibitemOpen
  \bibfield  {author} {\bibinfo {author} {\bibfnamefont {T.~E.}\ \bibnamefont
  {Golikova}}, \bibinfo {author} {\bibfnamefont {M.~J.}\ \bibnamefont {Wolf}},
  \bibinfo {author} {\bibfnamefont {D.}~\bibnamefont {Beckmann}}, \bibinfo
  {author} {\bibfnamefont {G.~A.}\ \bibnamefont {Penzyakov}}, \bibinfo {author}
  {\bibfnamefont {I.~E.}\ \bibnamefont {Batov}}, \bibinfo {author}
  {\bibfnamefont {I.}~\bibnamefont {Bobkova}}, \bibinfo {author} {\bibfnamefont
  {A.~M.}\ \bibnamefont {Bobkov}},\ and\ \bibinfo {author} {\bibfnamefont
  {V.~V.}\ \bibnamefont {Ryazanov}},\ }\href
  {https://iopscience.iop.org/article/10.1088/1361-6668/abfd0d/meta} {\bibfield
   {journal} {\bibinfo  {journal} {Superconductor Science and Technology}\ }
  (\bibinfo {year} {2021})}\BibitemShut {NoStop}%
\bibitem [{\citenamefont {van Dam}\ \emph {et~al.}(2006)\citenamefont {van
  Dam}, \citenamefont {Nazarov}, \citenamefont {Bakkers}, \citenamefont
  {De~Franceschi},\ and\ \citenamefont {Kouwenhoven}}]{vanDam2006}%
  \BibitemOpen
  \bibfield  {author} {\bibinfo {author} {\bibfnamefont {J.~A.}\ \bibnamefont
  {van Dam}}, \bibinfo {author} {\bibfnamefont {Y.~V.}\ \bibnamefont
  {Nazarov}}, \bibinfo {author} {\bibfnamefont {E.~P. A.~M.}\ \bibnamefont
  {Bakkers}}, \bibinfo {author} {\bibfnamefont {S.}~\bibnamefont
  {De~Franceschi}},\ and\ \bibinfo {author} {\bibfnamefont {L.~P.}\
  \bibnamefont {Kouwenhoven}},\ }\href {https://doi.org/10.1038/nature05018}
  {\bibfield  {journal} {\bibinfo  {journal} {Nature}\ }\textbf {\bibinfo
  {volume} {442}},\ \bibinfo {pages} {667} (\bibinfo {year}
  {2006})}\BibitemShut {NoStop}%
\bibitem [{\citenamefont {J{\o}rgensen}\ \emph {et~al.}(2007)\citenamefont
  {J{\o}rgensen}, \citenamefont {Novotn{\`y}}, \citenamefont {Grove-Rasmussen},
  \citenamefont {Flensberg},\ and\ \citenamefont
  {Lindelof}}]{jorgensen2007critical}%
  \BibitemOpen
  \bibfield  {author} {\bibinfo {author} {\bibfnamefont {H.~I.}\ \bibnamefont
  {J{\o}rgensen}}, \bibinfo {author} {\bibfnamefont {T.}~\bibnamefont
  {Novotn{\`y}}}, \bibinfo {author} {\bibfnamefont {K.}~\bibnamefont
  {Grove-Rasmussen}}, \bibinfo {author} {\bibfnamefont {K.}~\bibnamefont
  {Flensberg}},\ and\ \bibinfo {author} {\bibfnamefont {P.}~\bibnamefont
  {Lindelof}},\ }\href {https://pubs.acs.org/doi/10.1021/nl071152w} {\bibfield
  {journal} {\bibinfo  {journal} {Nano letters}\ }\textbf {\bibinfo {volume}
  {7}},\ \bibinfo {pages} {2441} (\bibinfo {year} {2007})}\BibitemShut
  {NoStop}%
\bibitem [{\citenamefont {Ke}\ \emph {et~al.}(2019)\citenamefont {Ke},
  \citenamefont {Moehle}, \citenamefont {de~Vries}, \citenamefont {Thomas},
  \citenamefont {Metti}, \citenamefont {Guinn}, \citenamefont {Kallaher},
  \citenamefont {Lodari}, \citenamefont {Scappucci}, \citenamefont {Wang} \emph
  {et~al.}}]{ke2019ballistic}%
  \BibitemOpen
  \bibfield  {author} {\bibinfo {author} {\bibfnamefont {C.~T.}\ \bibnamefont
  {Ke}}, \bibinfo {author} {\bibfnamefont {C.~M.}\ \bibnamefont {Moehle}},
  \bibinfo {author} {\bibfnamefont {F.~K.}\ \bibnamefont {de~Vries}}, \bibinfo
  {author} {\bibfnamefont {C.}~\bibnamefont {Thomas}}, \bibinfo {author}
  {\bibfnamefont {S.}~\bibnamefont {Metti}}, \bibinfo {author} {\bibfnamefont
  {C.~R.}\ \bibnamefont {Guinn}}, \bibinfo {author} {\bibfnamefont
  {R.}~\bibnamefont {Kallaher}}, \bibinfo {author} {\bibfnamefont
  {M.}~\bibnamefont {Lodari}}, \bibinfo {author} {\bibfnamefont
  {G.}~\bibnamefont {Scappucci}}, \bibinfo {author} {\bibfnamefont
  {T.}~\bibnamefont {Wang}}, \emph {et~al.},\ }\href
  {https://www.nature.com/articles/s41467-019-11742-4} {\bibfield  {journal}
  {\bibinfo  {journal} {Nature communications}\ }\textbf {\bibinfo {volume}
  {10}},\ \bibinfo {pages} {1} (\bibinfo {year} {2019})}\BibitemShut {NoStop}%
\bibitem [{\citenamefont {Volkov}(1995)}]{Volkov1995}%
  \BibitemOpen
  \bibfield  {author} {\bibinfo {author} {\bibfnamefont {A.~F.}\ \bibnamefont
  {Volkov}},\ }\href {https://doi.org/10.1103/PhysRevLett.74.4730} {\bibfield
  {journal} {\bibinfo  {journal} {Phys. Rev. Lett.}\ }\textbf {\bibinfo
  {volume} {74}},\ \bibinfo {pages} {4730} (\bibinfo {year}
  {1995})}\BibitemShut {NoStop}%
\bibitem [{\citenamefont {Morpurgo}\ \emph {et~al.}(1998)\citenamefont
  {Morpurgo}, \citenamefont {Klapwijk},\ and\ \citenamefont {van
  Wees}}]{Morpurgo1998}%
  \BibitemOpen
  \bibfield  {author} {\bibinfo {author} {\bibfnamefont {A.~F.}\ \bibnamefont
  {Morpurgo}}, \bibinfo {author} {\bibfnamefont {T.~M.}\ \bibnamefont
  {Klapwijk}},\ and\ \bibinfo {author} {\bibfnamefont {B.~J.}\ \bibnamefont
  {van Wees}},\ }\href {https://doi.org/10.1063/1.120612} {\bibfield  {journal}
  {\bibinfo  {journal} {Applied Physics Letters}\ }\textbf {\bibinfo {volume}
  {72}},\ \bibinfo {pages} {966} (\bibinfo {year} {1998})}\BibitemShut
  {NoStop}%
\bibitem [{\citenamefont {Huang}\ \emph {et~al.}(2002)\citenamefont {Huang},
  \citenamefont {Pierre}, \citenamefont {Heikkil\"a}, \citenamefont {Wilhelm},\
  and\ \citenamefont {Birge}}]{Huang2002}%
  \BibitemOpen
  \bibfield  {author} {\bibinfo {author} {\bibfnamefont {J.}~\bibnamefont
  {Huang}}, \bibinfo {author} {\bibfnamefont {F.}~\bibnamefont {Pierre}},
  \bibinfo {author} {\bibfnamefont {T.~T.}\ \bibnamefont {Heikkil\"a}},
  \bibinfo {author} {\bibfnamefont {F.~K.}\ \bibnamefont {Wilhelm}},\ and\
  \bibinfo {author} {\bibfnamefont {N.~O.}\ \bibnamefont {Birge}},\ }\href
  {https://doi.org/10.1103/PhysRevB.66.020507} {\bibfield  {journal} {\bibinfo
  {journal} {Phys. Rev. B}\ }\textbf {\bibinfo {volume} {66}},\ \bibinfo
  {pages} {020507(R)} (\bibinfo {year} {2002})}\BibitemShut {NoStop}%
\bibitem [{\citenamefont {Yip}(2000)}]{Yip2000}%
  \BibitemOpen
  \bibfield  {author} {\bibinfo {author} {\bibfnamefont {S.-K.}\ \bibnamefont
  {Yip}},\ }\href {https://doi.org/10.1103/PhysRevB.62.R6127} {\bibfield
  {journal} {\bibinfo  {journal} {Phys. Rev. B}\ }\textbf {\bibinfo {volume}
  {62}},\ \bibinfo {pages} {R6127} (\bibinfo {year} {2000})}\BibitemShut
  {NoStop}%
\bibitem [{\citenamefont {Heikkil{\"a}}\ \emph {et~al.}(2000)\citenamefont
  {Heikkil{\"a}}, \citenamefont {Wilhelm},\ and\ \citenamefont
  {Sch{\"o}n}}]{Heikkila2000}%
  \BibitemOpen
  \bibfield  {author} {\bibinfo {author} {\bibfnamefont {T.~T.}\ \bibnamefont
  {Heikkil{\"a}}}, \bibinfo {author} {\bibfnamefont {F.~K.}\ \bibnamefont
  {Wilhelm}},\ and\ \bibinfo {author} {\bibfnamefont {G.}~\bibnamefont
  {Sch{\"o}n}},\ }\href {https://doi.org/10.1209/epl/i2000-00513-x} {\bibfield
  {journal} {\bibinfo  {journal} {Europhysics Letters (EPL)}\ }\textbf
  {\bibinfo {volume} {51}},\ \bibinfo {pages} {434} (\bibinfo {year}
  {2000})}\BibitemShut {NoStop}%
\bibitem [{\citenamefont {Wilhelm}\ \emph {et~al.}(1998)\citenamefont
  {Wilhelm}, \citenamefont {Sch\"on},\ and\ \citenamefont
  {Zaikin}}]{Wilheim1998}%
  \BibitemOpen
  \bibfield  {author} {\bibinfo {author} {\bibfnamefont {F.~K.}\ \bibnamefont
  {Wilhelm}}, \bibinfo {author} {\bibfnamefont {G.}~\bibnamefont {Sch\"on}},\
  and\ \bibinfo {author} {\bibfnamefont {A.~D.}\ \bibnamefont {Zaikin}},\
  }\href {https://doi.org/10.1103/PhysRevLett.81.1682} {\bibfield  {journal}
  {\bibinfo  {journal} {Phys. Rev. Lett.}\ }\textbf {\bibinfo {volume} {81}},\
  \bibinfo {pages} {1682} (\bibinfo {year} {1998})}\BibitemShut {NoStop}%
\bibitem [{\citenamefont {Bobkova}\ and\ \citenamefont
  {Bobkov}(2010)}]{Bobkova2010}%
  \BibitemOpen
  \bibfield  {author} {\bibinfo {author} {\bibfnamefont {I.~V.}\ \bibnamefont
  {Bobkova}}\ and\ \bibinfo {author} {\bibfnamefont {A.~M.}\ \bibnamefont
  {Bobkov}},\ }\href {https://doi.org/10.1103/PhysRevB.82.024515} {\bibfield
  {journal} {\bibinfo  {journal} {Phys. Rev. B}\ }\textbf {\bibinfo {volume}
  {82}},\ \bibinfo {pages} {024515} (\bibinfo {year} {2010})}\BibitemShut
  {NoStop}%
\bibitem [{\citenamefont {Bobkov}\ and\ \citenamefont
  {Bobkova}(2011)}]{Bobkov2011}%
  \BibitemOpen
  \bibfield  {author} {\bibinfo {author} {\bibfnamefont {A.~M.}\ \bibnamefont
  {Bobkov}}\ and\ \bibinfo {author} {\bibfnamefont {I.~V.}\ \bibnamefont
  {Bobkova}},\ }\href {https://doi.org/10.1103/PhysRevB.84.054533} {\bibfield
  {journal} {\bibinfo  {journal} {Phys. Rev. B}\ }\textbf {\bibinfo {volume}
  {84}},\ \bibinfo {pages} {054533} (\bibinfo {year} {2011})}\BibitemShut
  {NoStop}%
\bibitem [{\citenamefont {Gingrich}\ \emph {et~al.}(2016)\citenamefont
  {Gingrich}, \citenamefont {Niedzielski}, \citenamefont {Glick}, \citenamefont
  {Wang}, \citenamefont {Miller}, \citenamefont {Loloee}, \citenamefont
  {Pratt~Jr},\ and\ \citenamefont {Birge}}]{Gingrich2016}%
  \BibitemOpen
  \bibfield  {author} {\bibinfo {author} {\bibfnamefont {E.~C.}\ \bibnamefont
  {Gingrich}}, \bibinfo {author} {\bibfnamefont {B.~M.}\ \bibnamefont
  {Niedzielski}}, \bibinfo {author} {\bibfnamefont {J.~A.}\ \bibnamefont
  {Glick}}, \bibinfo {author} {\bibfnamefont {Y.}~\bibnamefont {Wang}},
  \bibinfo {author} {\bibfnamefont {D.~L.}\ \bibnamefont {Miller}}, \bibinfo
  {author} {\bibfnamefont {R.}~\bibnamefont {Loloee}}, \bibinfo {author}
  {\bibfnamefont {W.~P.}\ \bibnamefont {Pratt~Jr}},\ and\ \bibinfo {author}
  {\bibfnamefont {N.~O.}\ \bibnamefont {Birge}},\ }\href
  {https://doi.org/10.1038/nphys3681} {\bibfield  {journal} {\bibinfo
  {journal} {Nature Physics}\ }\textbf {\bibinfo {volume} {12}},\ \bibinfo
  {pages} {564} (\bibinfo {year} {2016})}\BibitemShut {NoStop}%
\bibitem [{\citenamefont {Bujnowski}\ \emph {et~al.}(2019)\citenamefont
  {Bujnowski}, \citenamefont {Biele},\ and\ \citenamefont
  {Bergeret}}]{Bujnowski2019}%
  \BibitemOpen
  \bibfield  {author} {\bibinfo {author} {\bibfnamefont {B.}~\bibnamefont
  {Bujnowski}}, \bibinfo {author} {\bibfnamefont {R.}~\bibnamefont {Biele}},\
  and\ \bibinfo {author} {\bibfnamefont {F.~S.}\ \bibnamefont {Bergeret}},\
  }\href {https://doi.org/10.1103/PhysRevB.100.224518} {\bibfield  {journal}
  {\bibinfo  {journal} {Phys. Rev. B}\ }\textbf {\bibinfo {volume} {100}},\
  \bibinfo {pages} {224518} (\bibinfo {year} {2019})}\BibitemShut {NoStop}%
\bibitem [{\citenamefont {Eskilt}\ \emph {et~al.}(2019)\citenamefont {Eskilt},
  \citenamefont {Amundsen}, \citenamefont {Banerjee},\ and\ \citenamefont
  {Linder}}]{Eskilt2019}%
  \BibitemOpen
  \bibfield  {author} {\bibinfo {author} {\bibfnamefont {J.~R.}\ \bibnamefont
  {Eskilt}}, \bibinfo {author} {\bibfnamefont {M.}~\bibnamefont {Amundsen}},
  \bibinfo {author} {\bibfnamefont {N.}~\bibnamefont {Banerjee}},\ and\
  \bibinfo {author} {\bibfnamefont {J.}~\bibnamefont {Linder}},\ }\href
  {https://doi.org/10.1103/PhysRevB.100.224519} {\bibfield  {journal} {\bibinfo
   {journal} {Phys. Rev. B}\ }\textbf {\bibinfo {volume} {100}},\ \bibinfo
  {pages} {224519} (\bibinfo {year} {2019})}\BibitemShut {NoStop}%
\bibitem [{\citenamefont {Paolucci}\ \emph {et~al.}(2019)\citenamefont
  {Paolucci}, \citenamefont {Vischi}, \citenamefont {De~Simoni}, \citenamefont
  {Guarcello}, \citenamefont {Solinas},\ and\ \citenamefont
  {Giazotto}}]{Paolucci2019}%
  \BibitemOpen
  \bibfield  {author} {\bibinfo {author} {\bibfnamefont {F.}~\bibnamefont
  {Paolucci}}, \bibinfo {author} {\bibfnamefont {F.}~\bibnamefont {Vischi}},
  \bibinfo {author} {\bibfnamefont {G.}~\bibnamefont {De~Simoni}}, \bibinfo
  {author} {\bibfnamefont {C.}~\bibnamefont {Guarcello}}, \bibinfo {author}
  {\bibfnamefont {P.}~\bibnamefont {Solinas}},\ and\ \bibinfo {author}
  {\bibfnamefont {F.}~\bibnamefont {Giazotto}},\ }\href
  {https://doi.org/10.1021/acs.nanolett.9b02369} {\bibfield  {journal}
  {\bibinfo  {journal} {Nano Letters}\ }\textbf {\bibinfo {volume} {19}},\
  \bibinfo {pages} {6263} (\bibinfo {year} {2019})}\BibitemShut {NoStop}%
\bibitem [{\citenamefont {De~Simoni}\ \emph {et~al.}(2018)\citenamefont
  {De~Simoni}, \citenamefont {Paolucci}, \citenamefont {Solinas}, \citenamefont
  {Strambini},\ and\ \citenamefont {Giazotto}}]{DeSimoni2018}%
  \BibitemOpen
  \bibfield  {author} {\bibinfo {author} {\bibfnamefont {G.}~\bibnamefont
  {De~Simoni}}, \bibinfo {author} {\bibfnamefont {F.}~\bibnamefont {Paolucci}},
  \bibinfo {author} {\bibfnamefont {P.}~\bibnamefont {Solinas}}, \bibinfo
  {author} {\bibfnamefont {E.}~\bibnamefont {Strambini}},\ and\ \bibinfo
  {author} {\bibfnamefont {F.}~\bibnamefont {Giazotto}},\ }\href
  {https://doi.org/10.1038/s41565-018-0190-3} {\bibfield  {journal} {\bibinfo
  {journal} {Nature Nanotechnology}\ }\textbf {\bibinfo {volume} {13}},\
  \bibinfo {pages} {802} (\bibinfo {year} {2018})}\BibitemShut {NoStop}%
\bibitem [{\citenamefont {Larsen}\ \emph {et~al.}(2015)\citenamefont {Larsen},
  \citenamefont {Petersson}, \citenamefont {Kuemmeth}, \citenamefont
  {Jespersen}, \citenamefont {Krogstrup}, \citenamefont {Nyg\aa{}rd},\ and\
  \citenamefont {Marcus}}]{Larsen2015}%
  \BibitemOpen
  \bibfield  {author} {\bibinfo {author} {\bibfnamefont {T.~W.}\ \bibnamefont
  {Larsen}}, \bibinfo {author} {\bibfnamefont {K.~D.}\ \bibnamefont
  {Petersson}}, \bibinfo {author} {\bibfnamefont {F.}~\bibnamefont {Kuemmeth}},
  \bibinfo {author} {\bibfnamefont {T.~S.}\ \bibnamefont {Jespersen}}, \bibinfo
  {author} {\bibfnamefont {P.}~\bibnamefont {Krogstrup}}, \bibinfo {author}
  {\bibfnamefont {J.}~\bibnamefont {Nyg\aa{}rd}},\ and\ \bibinfo {author}
  {\bibfnamefont {C.~M.}\ \bibnamefont {Marcus}},\ }\href
  {https://doi.org/10.1103/PhysRevLett.115.127001} {\bibfield  {journal}
  {\bibinfo  {journal} {Phys. Rev. Lett.}\ }\textbf {\bibinfo {volume} {115}},\
  \bibinfo {pages} {127001} (\bibinfo {year} {2015})}\BibitemShut {NoStop}%
\bibitem [{\citenamefont {Casparis}\ \emph {et~al.}(2016)\citenamefont
  {Casparis}, \citenamefont {Larsen}, \citenamefont {Olsen}, \citenamefont
  {Kuemmeth}, \citenamefont {Krogstrup}, \citenamefont {Nyg\aa{}rd},
  \citenamefont {Petersson},\ and\ \citenamefont {Marcus}}]{Casparis2016}%
  \BibitemOpen
  \bibfield  {author} {\bibinfo {author} {\bibfnamefont {L.}~\bibnamefont
  {Casparis}}, \bibinfo {author} {\bibfnamefont {T.~W.}\ \bibnamefont
  {Larsen}}, \bibinfo {author} {\bibfnamefont {M.~S.}\ \bibnamefont {Olsen}},
  \bibinfo {author} {\bibfnamefont {F.}~\bibnamefont {Kuemmeth}}, \bibinfo
  {author} {\bibfnamefont {P.}~\bibnamefont {Krogstrup}}, \bibinfo {author}
  {\bibfnamefont {J.}~\bibnamefont {Nyg\aa{}rd}}, \bibinfo {author}
  {\bibfnamefont {K.~D.}\ \bibnamefont {Petersson}},\ and\ \bibinfo {author}
  {\bibfnamefont {C.~M.}\ \bibnamefont {Marcus}},\ }\href
  {https://doi.org/10.1103/PhysRevLett.116.150505} {\bibfield  {journal}
  {\bibinfo  {journal} {Phys. Rev. Lett.}\ }\textbf {\bibinfo {volume} {116}},\
  \bibinfo {pages} {150505} (\bibinfo {year} {2016})}\BibitemShut {NoStop}%
\bibitem [{\citenamefont {Doh}\ \emph {et~al.}(2005)\citenamefont {Doh},
  \citenamefont {van Dam}, \citenamefont {Roest}, \citenamefont {Bakkers},
  \citenamefont {Kouwenhoven},\ and\ \citenamefont {De~Franceschi}}]{Doh2005}%
  \BibitemOpen
  \bibfield  {author} {\bibinfo {author} {\bibfnamefont {Y.-J.}\ \bibnamefont
  {Doh}}, \bibinfo {author} {\bibfnamefont {J.~A.}\ \bibnamefont {van Dam}},
  \bibinfo {author} {\bibfnamefont {A.~L.}\ \bibnamefont {Roest}}, \bibinfo
  {author} {\bibfnamefont {E.~P. A.~M.}\ \bibnamefont {Bakkers}}, \bibinfo
  {author} {\bibfnamefont {L.~P.}\ \bibnamefont {Kouwenhoven}},\ and\ \bibinfo
  {author} {\bibfnamefont {S.}~\bibnamefont {De~Franceschi}},\ }\href
  {https://doi.org/10.1126/science.1113523} {\bibfield  {journal} {\bibinfo
  {journal} {Science}\ }\textbf {\bibinfo {volume} {309}},\ \bibinfo {pages}
  {272} (\bibinfo {year} {2005})}\BibitemShut {NoStop}%
\bibitem [{\citenamefont {Abay}\ \emph {et~al.}(2014)\citenamefont {Abay},
  \citenamefont {Persson}, \citenamefont {Nilsson}, \citenamefont {Wu},
  \citenamefont {Xu}, \citenamefont {Fogelstr\"om}, \citenamefont {Shumeiko},\
  and\ \citenamefont {Delsing}}]{Abay2014}%
  \BibitemOpen
  \bibfield  {author} {\bibinfo {author} {\bibfnamefont {S.}~\bibnamefont
  {Abay}}, \bibinfo {author} {\bibfnamefont {D.}~\bibnamefont {Persson}},
  \bibinfo {author} {\bibfnamefont {H.}~\bibnamefont {Nilsson}}, \bibinfo
  {author} {\bibfnamefont {F.}~\bibnamefont {Wu}}, \bibinfo {author}
  {\bibfnamefont {H.~Q.}\ \bibnamefont {Xu}}, \bibinfo {author} {\bibfnamefont
  {M.}~\bibnamefont {Fogelstr\"om}}, \bibinfo {author} {\bibfnamefont
  {V.}~\bibnamefont {Shumeiko}},\ and\ \bibinfo {author} {\bibfnamefont
  {P.}~\bibnamefont {Delsing}},\ }\href
  {https://doi.org/10.1103/PhysRevB.89.214508} {\bibfield  {journal} {\bibinfo
  {journal} {Phys. Rev. B}\ }\textbf {\bibinfo {volume} {89}},\ \bibinfo
  {pages} {214508} (\bibinfo {year} {2014})}\BibitemShut {NoStop}%
\bibitem [{\citenamefont {Bobkova}\ and\ \citenamefont
  {Bobkov}(2012)}]{Bobkova2012}%
  \BibitemOpen
  \bibfield  {author} {\bibinfo {author} {\bibfnamefont {I.~V.}\ \bibnamefont
  {Bobkova}}\ and\ \bibinfo {author} {\bibfnamefont {A.~M.}\ \bibnamefont
  {Bobkov}},\ }\href {https://doi.org/10.1103/PhysRevLett.108.197002}
  {\bibfield  {journal} {\bibinfo  {journal} {Phys. Rev. Lett.}\ }\textbf
  {\bibinfo {volume} {108}},\ \bibinfo {pages} {197002} (\bibinfo {year}
  {2012})}\BibitemShut {NoStop}%
\bibitem [{\citenamefont {Bergeret}\ \emph
  {et~al.}(2001{\natexlab{a}})\citenamefont {Bergeret}, \citenamefont
  {Volkov},\ and\ \citenamefont {Efetov}}]{Bergeret2001}%
  \BibitemOpen
  \bibfield  {author} {\bibinfo {author} {\bibfnamefont {F.~S.}\ \bibnamefont
  {Bergeret}}, \bibinfo {author} {\bibfnamefont {A.~F.}\ \bibnamefont
  {Volkov}},\ and\ \bibinfo {author} {\bibfnamefont {K.~B.}\ \bibnamefont
  {Efetov}},\ }\href {https://doi.org/10.1103/PhysRevLett.86.4096} {\bibfield
  {journal} {\bibinfo  {journal} {Phys. Rev. Lett.}\ }\textbf {\bibinfo
  {volume} {86}},\ \bibinfo {pages} {4096} (\bibinfo {year}
  {2001}{\natexlab{a}})}\BibitemShut {NoStop}%
\bibitem [{\citenamefont {Bergeret}\ \emph
  {et~al.}(2001{\natexlab{b}})\citenamefont {Bergeret}, \citenamefont
  {Volkov},\ and\ \citenamefont {Efetov}}]{Bergeret2001_2}%
  \BibitemOpen
  \bibfield  {author} {\bibinfo {author} {\bibfnamefont {F.~S.}\ \bibnamefont
  {Bergeret}}, \bibinfo {author} {\bibfnamefont {A.~F.}\ \bibnamefont
  {Volkov}},\ and\ \bibinfo {author} {\bibfnamefont {K.~B.}\ \bibnamefont
  {Efetov}},\ }\href {https://doi.org/10.1103/PhysRevLett.86.3140} {\bibfield
  {journal} {\bibinfo  {journal} {Phys. Rev. Lett.}\ }\textbf {\bibinfo
  {volume} {86}},\ \bibinfo {pages} {3140} (\bibinfo {year}
  {2001}{\natexlab{b}})}\BibitemShut {NoStop}%
\bibitem [{\citenamefont {{Kadigrobov, A.}}\ \emph {et~al.}(2001)\citenamefont
  {{Kadigrobov, A.}}, \citenamefont {{Shekhter, R. I.}},\ and\ \citenamefont
  {{Jonson, M.}}}]{Kadigrobov2001}%
  \BibitemOpen
  \bibfield  {author} {\bibinfo {author} {\bibnamefont {{Kadigrobov, A.}}},
  \bibinfo {author} {\bibnamefont {{Shekhter, R. I.}}},\ and\ \bibinfo {author}
  {\bibnamefont {{Jonson, M.}}},\ }\href
  {https://doi.org/10.1209/epl/i2001-00107-2} {\bibfield  {journal} {\bibinfo
  {journal} {Europhys. Lett.}\ }\textbf {\bibinfo {volume} {54}},\ \bibinfo
  {pages} {394} (\bibinfo {year} {2001})}\BibitemShut {NoStop}%
\bibitem [{\citenamefont {Fominov}\ \emph {et~al.}(2003)\citenamefont
  {Fominov}, \citenamefont {Golubov},\ and\ \citenamefont
  {Kupriyanov}}]{Fominov2003}%
  \BibitemOpen
  \bibfield  {author} {\bibinfo {author} {\bibfnamefont {Y.~V.}\ \bibnamefont
  {Fominov}}, \bibinfo {author} {\bibfnamefont {A.~A.}\ \bibnamefont
  {Golubov}},\ and\ \bibinfo {author} {\bibfnamefont {M.~Y.}\ \bibnamefont
  {Kupriyanov}},\ }\href {https://doi.org/10.1134/1.1591981} {\bibfield
  {journal} {\bibinfo  {journal} {Journal of Experimental and Theoretical
  Physics Letters}\ }\textbf {\bibinfo {volume} {77}},\ \bibinfo {pages} {510}
  (\bibinfo {year} {2003})}\BibitemShut {NoStop}%
\bibitem [{\citenamefont {Bergeret}\ \emph {et~al.}(2005)\citenamefont
  {Bergeret}, \citenamefont {Volkov},\ and\ \citenamefont
  {Efetov}}]{Bergeret2005}%
  \BibitemOpen
  \bibfield  {author} {\bibinfo {author} {\bibfnamefont {F.~S.}\ \bibnamefont
  {Bergeret}}, \bibinfo {author} {\bibfnamefont {A.~F.}\ \bibnamefont
  {Volkov}},\ and\ \bibinfo {author} {\bibfnamefont {K.~B.}\ \bibnamefont
  {Efetov}},\ }\href {https://doi.org/10.1103/RevModPhys.77.1321} {\bibfield
  {journal} {\bibinfo  {journal} {Rev. Mod. Phys.}\ }\textbf {\bibinfo {volume}
  {77}},\ \bibinfo {pages} {1321} (\bibinfo {year} {2005})}\BibitemShut
  {NoStop}%
\bibitem [{\citenamefont {Houzet}\ and\ \citenamefont
  {Buzdin}(2007)}]{Houzet2007}%
  \BibitemOpen
  \bibfield  {author} {\bibinfo {author} {\bibfnamefont {M.}~\bibnamefont
  {Houzet}}\ and\ \bibinfo {author} {\bibfnamefont {A.~I.}\ \bibnamefont
  {Buzdin}},\ }\href {https://doi.org/10.1103/PhysRevB.76.060504} {\bibfield
  {journal} {\bibinfo  {journal} {Phys. Rev. B}\ }\textbf {\bibinfo {volume}
  {76}},\ \bibinfo {pages} {060504(R)} (\bibinfo {year} {2007})}\BibitemShut
  {NoStop}%
\bibitem [{\citenamefont {Fominov}\ \emph {et~al.}(2007)\citenamefont
  {Fominov}, \citenamefont {Volkov},\ and\ \citenamefont
  {Efetov}}]{fominov2007josephson}%
  \BibitemOpen
  \bibfield  {author} {\bibinfo {author} {\bibfnamefont {Y.~V.}\ \bibnamefont
  {Fominov}}, \bibinfo {author} {\bibfnamefont {A.~F.}\ \bibnamefont
  {Volkov}},\ and\ \bibinfo {author} {\bibfnamefont {K.~B.}\ \bibnamefont
  {Efetov}},\ }\href
  {https://journals.aps.org/prb/abstract/10.1103/PhysRevB.75.104509} {\bibfield
   {journal} {\bibinfo  {journal} {Physical Review B}\ }\textbf {\bibinfo
  {volume} {75}},\ \bibinfo {pages} {104509} (\bibinfo {year}
  {2007})}\BibitemShut {NoStop}%
\bibitem [{\citenamefont {Fominov}\ \emph {et~al.}(2010)\citenamefont
  {Fominov}, \citenamefont {Golubov}, \citenamefont {Karminskaya},
  \citenamefont {Kupriyanov}, \citenamefont {Deminov},\ and\ \citenamefont
  {Tagirov}}]{Fominov2010}%
  \BibitemOpen
  \bibfield  {author} {\bibinfo {author} {\bibfnamefont {Y.~V.}\ \bibnamefont
  {Fominov}}, \bibinfo {author} {\bibfnamefont {A.~A.}\ \bibnamefont
  {Golubov}}, \bibinfo {author} {\bibfnamefont {T.~Y.}\ \bibnamefont
  {Karminskaya}}, \bibinfo {author} {\bibfnamefont {M.~Y.}\ \bibnamefont
  {Kupriyanov}}, \bibinfo {author} {\bibfnamefont {R.~G.}\ \bibnamefont
  {Deminov}},\ and\ \bibinfo {author} {\bibfnamefont {L.~R.}\ \bibnamefont
  {Tagirov}},\ }\href {https://doi.org/10.1134/S002136401006010X} {\bibfield
  {journal} {\bibinfo  {journal} {JETP Letters}\ }\textbf {\bibinfo {volume}
  {91}},\ \bibinfo {pages} {308} (\bibinfo {year} {2010})}\BibitemShut
  {NoStop}%
\bibitem [{\citenamefont {Halterman}\ \emph {et~al.}(2007)\citenamefont
  {Halterman}, \citenamefont {Barsic},\ and\ \citenamefont
  {Valls}}]{Halterman2007}%
  \BibitemOpen
  \bibfield  {author} {\bibinfo {author} {\bibfnamefont {K.}~\bibnamefont
  {Halterman}}, \bibinfo {author} {\bibfnamefont {P.~H.}\ \bibnamefont
  {Barsic}},\ and\ \bibinfo {author} {\bibfnamefont {O.~T.}\ \bibnamefont
  {Valls}},\ }\href {https://doi.org/10.1103/PhysRevLett.99.127002} {\bibfield
  {journal} {\bibinfo  {journal} {Phys. Rev. Lett.}\ }\textbf {\bibinfo
  {volume} {99}},\ \bibinfo {pages} {127002} (\bibinfo {year}
  {2007})}\BibitemShut {NoStop}%
\bibitem [{\citenamefont {Halterman}\ \emph {et~al.}(2008)\citenamefont
  {Halterman}, \citenamefont {Valls},\ and\ \citenamefont
  {Barsic}}]{Halterman2008}%
  \BibitemOpen
  \bibfield  {author} {\bibinfo {author} {\bibfnamefont {K.}~\bibnamefont
  {Halterman}}, \bibinfo {author} {\bibfnamefont {O.~T.}\ \bibnamefont
  {Valls}},\ and\ \bibinfo {author} {\bibfnamefont {P.~H.}\ \bibnamefont
  {Barsic}},\ }\href {https://doi.org/10.1103/PhysRevB.77.174511} {\bibfield
  {journal} {\bibinfo  {journal} {Phys. Rev. B}\ }\textbf {\bibinfo {volume}
  {77}},\ \bibinfo {pages} {174511} (\bibinfo {year} {2008})}\BibitemShut
  {NoStop}%
\bibitem [{\citenamefont {Zhu}\ \emph {et~al.}(2010)\citenamefont {Zhu},
  \citenamefont {Krivorotov}, \citenamefont {Halterman},\ and\ \citenamefont
  {Valls}}]{Zhu2010}%
  \BibitemOpen
  \bibfield  {author} {\bibinfo {author} {\bibfnamefont {J.}~\bibnamefont
  {Zhu}}, \bibinfo {author} {\bibfnamefont {I.~N.}\ \bibnamefont {Krivorotov}},
  \bibinfo {author} {\bibfnamefont {K.}~\bibnamefont {Halterman}},\ and\
  \bibinfo {author} {\bibfnamefont {O.~T.}\ \bibnamefont {Valls}},\ }\href
  {https://doi.org/10.1103/PhysRevLett.105.207002} {\bibfield  {journal}
  {\bibinfo  {journal} {Phys. Rev. Lett.}\ }\textbf {\bibinfo {volume} {105}},\
  \bibinfo {pages} {207002} (\bibinfo {year} {2010})}\BibitemShut {NoStop}%
\bibitem [{\citenamefont {Alidoust}\ \emph {et~al.}(2014)\citenamefont
  {Alidoust}, \citenamefont {Halterman},\ and\ \citenamefont
  {Linder}}]{Alidoust2014}%
  \BibitemOpen
  \bibfield  {author} {\bibinfo {author} {\bibfnamefont {M.}~\bibnamefont
  {Alidoust}}, \bibinfo {author} {\bibfnamefont {K.}~\bibnamefont
  {Halterman}},\ and\ \bibinfo {author} {\bibfnamefont {J.}~\bibnamefont
  {Linder}},\ }\href {https://doi.org/10.1103/PhysRevB.89.054508} {\bibfield
  {journal} {\bibinfo  {journal} {Phys. Rev. B}\ }\textbf {\bibinfo {volume}
  {89}},\ \bibinfo {pages} {054508} (\bibinfo {year} {2014})}\BibitemShut
  {NoStop}%
\bibitem [{\citenamefont {Keizer}\ \emph {et~al.}(2006)\citenamefont {Keizer},
  \citenamefont {Goennenwein}, \citenamefont {Klapwijk}, \citenamefont {Miao},
  \citenamefont {Xiao},\ and\ \citenamefont {Gupta}}]{Keizer2006}%
  \BibitemOpen
  \bibfield  {author} {\bibinfo {author} {\bibfnamefont {R.}~\bibnamefont
  {Keizer}}, \bibinfo {author} {\bibfnamefont {S.}~\bibnamefont {Goennenwein}},
  \bibinfo {author} {\bibfnamefont {T.}~\bibnamefont {Klapwijk}}, \bibinfo
  {author} {\bibfnamefont {G.}~\bibnamefont {Miao}}, \bibinfo {author}
  {\bibfnamefont {G.}~\bibnamefont {Xiao}},\ and\ \bibinfo {author}
  {\bibfnamefont {A.}~\bibnamefont {Gupta}},\ }\href
  {https://www.nature.com/articles/nature04499} {\bibfield  {journal} {\bibinfo
   {journal} {Nature}\ }\textbf {\bibinfo {volume} {439}},\ \bibinfo {pages}
  {825} (\bibinfo {year} {2006})}\BibitemShut {NoStop}%
\bibitem [{\citenamefont {Eschrig}\ \emph {et~al.}(2003)\citenamefont
  {Eschrig}, \citenamefont {Kopu}, \citenamefont {Cuevas},\ and\ \citenamefont
  {Sch\"on}}]{Eschrig2003}%
  \BibitemOpen
  \bibfield  {author} {\bibinfo {author} {\bibfnamefont {M.}~\bibnamefont
  {Eschrig}}, \bibinfo {author} {\bibfnamefont {J.}~\bibnamefont {Kopu}},
  \bibinfo {author} {\bibfnamefont {J.~C.}\ \bibnamefont {Cuevas}},\ and\
  \bibinfo {author} {\bibfnamefont {G.}~\bibnamefont {Sch\"on}},\ }\href
  {https://doi.org/10.1103/PhysRevLett.90.137003} {\bibfield  {journal}
  {\bibinfo  {journal} {Phys. Rev. Lett.}\ }\textbf {\bibinfo {volume} {90}},\
  \bibinfo {pages} {137003} (\bibinfo {year} {2003})}\BibitemShut {NoStop}%
\bibitem [{\citenamefont {Eschrig}\ and\ \citenamefont
  {L{\"o}fwander}(2008)}]{Eschrig2008}%
  \BibitemOpen
  \bibfield  {author} {\bibinfo {author} {\bibfnamefont {M.}~\bibnamefont
  {Eschrig}}\ and\ \bibinfo {author} {\bibfnamefont {T.}~\bibnamefont
  {L{\"o}fwander}},\ }\href {https://doi.org/10.1038/nphys831} {\bibfield
  {journal} {\bibinfo  {journal} {Nature Physics}\ }\textbf {\bibinfo {volume}
  {4}},\ \bibinfo {pages} {138} (\bibinfo {year} {2008})}\BibitemShut {NoStop}%
\bibitem [{\citenamefont {Robinson}\ \emph
  {et~al.}(2010{\natexlab{a}})\citenamefont {Robinson}, \citenamefont
  {Hal\'asz}, \citenamefont {Buzdin},\ and\ \citenamefont
  {Blamire}}]{Robinson2010}%
  \BibitemOpen
  \bibfield  {author} {\bibinfo {author} {\bibfnamefont {J.~W.~A.}\
  \bibnamefont {Robinson}}, \bibinfo {author} {\bibfnamefont {G.~B.}\
  \bibnamefont {Hal\'asz}}, \bibinfo {author} {\bibfnamefont {A.~I.}\
  \bibnamefont {Buzdin}},\ and\ \bibinfo {author} {\bibfnamefont {M.~G.}\
  \bibnamefont {Blamire}},\ }\href
  {https://doi.org/10.1103/PhysRevLett.104.207001} {\bibfield  {journal}
  {\bibinfo  {journal} {Phys. Rev. Lett.}\ }\textbf {\bibinfo {volume} {104}},\
  \bibinfo {pages} {207001} (\bibinfo {year} {2010}{\natexlab{a}})}\BibitemShut
  {NoStop}%
\bibitem [{\citenamefont {Braude}\ and\ \citenamefont
  {Nazarov}(2007)}]{Braude2007}%
  \BibitemOpen
  \bibfield  {author} {\bibinfo {author} {\bibfnamefont {V.}~\bibnamefont
  {Braude}}\ and\ \bibinfo {author} {\bibfnamefont {Y.~V.}\ \bibnamefont
  {Nazarov}},\ }\href {https://doi.org/10.1103/PhysRevLett.98.077003}
  {\bibfield  {journal} {\bibinfo  {journal} {Phys. Rev. Lett.}\ }\textbf
  {\bibinfo {volume} {98}},\ \bibinfo {pages} {077003} (\bibinfo {year}
  {2007})}\BibitemShut {NoStop}%
\bibitem [{\citenamefont {Robinson}\ \emph
  {et~al.}(2010{\natexlab{b}})\citenamefont {Robinson}, \citenamefont {Witt},\
  and\ \citenamefont {Blamire}}]{Robinson2010_2}%
  \BibitemOpen
  \bibfield  {author} {\bibinfo {author} {\bibfnamefont {J.}~\bibnamefont
  {Robinson}}, \bibinfo {author} {\bibfnamefont {J.}~\bibnamefont {Witt}},\
  and\ \bibinfo {author} {\bibfnamefont {M.}~\bibnamefont {Blamire}},\ }\href
  {https://www.science.org/doi/full/10.1126/science.1189246} {\bibfield
  {journal} {\bibinfo  {journal} {Science}\ }\textbf {\bibinfo {volume}
  {329}},\ \bibinfo {pages} {59} (\bibinfo {year}
  {2010}{\natexlab{b}})}\BibitemShut {NoStop}%
\bibitem [{\citenamefont {Singh}\ \emph {et~al.}(2016)\citenamefont {Singh},
  \citenamefont {Jansen}, \citenamefont {Lahabi},\ and\ \citenamefont
  {Aarts}}]{Singh2016}%
  \BibitemOpen
  \bibfield  {author} {\bibinfo {author} {\bibfnamefont {A.}~\bibnamefont
  {Singh}}, \bibinfo {author} {\bibfnamefont {C.}~\bibnamefont {Jansen}},
  \bibinfo {author} {\bibfnamefont {K.}~\bibnamefont {Lahabi}},\ and\ \bibinfo
  {author} {\bibfnamefont {J.}~\bibnamefont {Aarts}},\ }\href
  {https://doi.org/10.1103/PhysRevX.6.041012} {\bibfield  {journal} {\bibinfo
  {journal} {Phys. Rev. X}\ }\textbf {\bibinfo {volume} {6}},\ \bibinfo {pages}
  {041012} (\bibinfo {year} {2016})}\BibitemShut {NoStop}%
\bibitem [{\citenamefont {Khaire}\ \emph {et~al.}(2010)\citenamefont {Khaire},
  \citenamefont {Khasawneh}, \citenamefont {Pratt},\ and\ \citenamefont
  {Birge}}]{Khaire2010}%
  \BibitemOpen
  \bibfield  {author} {\bibinfo {author} {\bibfnamefont {T.~S.}\ \bibnamefont
  {Khaire}}, \bibinfo {author} {\bibfnamefont {M.~A.}\ \bibnamefont
  {Khasawneh}}, \bibinfo {author} {\bibfnamefont {W.~P.}\ \bibnamefont
  {Pratt}},\ and\ \bibinfo {author} {\bibfnamefont {N.~O.}\ \bibnamefont
  {Birge}},\ }\href {https://doi.org/10.1103/PhysRevLett.104.137002} {\bibfield
   {journal} {\bibinfo  {journal} {Phys. Rev. Lett.}\ }\textbf {\bibinfo
  {volume} {104}},\ \bibinfo {pages} {137002} (\bibinfo {year}
  {2010})}\BibitemShut {NoStop}%
\bibitem [{\citenamefont {Mironov}\ and\ \citenamefont
  {Buzdin}(2015)}]{Mironov2015}%
  \BibitemOpen
  \bibfield  {author} {\bibinfo {author} {\bibfnamefont {S.}~\bibnamefont
  {Mironov}}\ and\ \bibinfo {author} {\bibfnamefont {A.}~\bibnamefont
  {Buzdin}},\ }\href {https://doi.org/10.1103/PhysRevB.92.184506} {\bibfield
  {journal} {\bibinfo  {journal} {Phys. Rev. B}\ }\textbf {\bibinfo {volume}
  {92}},\ \bibinfo {pages} {184506} (\bibinfo {year} {2015})}\BibitemShut
  {NoStop}%
\bibitem [{\citenamefont {Halterman}\ and\ \citenamefont
  {Alidoust}(2016)}]{Halterman2016}%
  \BibitemOpen
  \bibfield  {author} {\bibinfo {author} {\bibfnamefont {K.}~\bibnamefont
  {Halterman}}\ and\ \bibinfo {author} {\bibfnamefont {M.}~\bibnamefont
  {Alidoust}},\ }\href {https://doi.org/10.1103/PhysRevB.94.064503} {\bibfield
  {journal} {\bibinfo  {journal} {Phys. Rev. B}\ }\textbf {\bibinfo {volume}
  {94}},\ \bibinfo {pages} {064503} (\bibinfo {year} {2016})}\BibitemShut
  {NoStop}%
\bibitem [{\citenamefont {Alidoust}\ and\ \citenamefont
  {Halterman}(2018)}]{Halterman2018}%
  \BibitemOpen
  \bibfield  {author} {\bibinfo {author} {\bibfnamefont {M.}~\bibnamefont
  {Alidoust}}\ and\ \bibinfo {author} {\bibfnamefont {K.}~\bibnamefont
  {Halterman}},\ }\href {https://doi.org/10.1103/PhysRevB.97.064517} {\bibfield
   {journal} {\bibinfo  {journal} {Phys. Rev. B}\ }\textbf {\bibinfo {volume}
  {97}},\ \bibinfo {pages} {064517} (\bibinfo {year} {2018})}\BibitemShut
  {NoStop}%
\bibitem [{\citenamefont {Srivastava}\ \emph {et~al.}(2017)\citenamefont
  {Srivastava}, \citenamefont {Olde~Olthof}, \citenamefont {Di~Bernardo},
  \citenamefont {Komori}, \citenamefont {Amado}, \citenamefont
  {Palomares-Garcia}, \citenamefont {Alidoust}, \citenamefont {Halterman},
  \citenamefont {Blamire},\ and\ \citenamefont {Robinson}}]{Srivastava2017}%
  \BibitemOpen
  \bibfield  {author} {\bibinfo {author} {\bibfnamefont {A.}~\bibnamefont
  {Srivastava}}, \bibinfo {author} {\bibfnamefont {L.~A.~B.}\ \bibnamefont
  {Olde~Olthof}}, \bibinfo {author} {\bibfnamefont {A.}~\bibnamefont
  {Di~Bernardo}}, \bibinfo {author} {\bibfnamefont {S.}~\bibnamefont {Komori}},
  \bibinfo {author} {\bibfnamefont {M.}~\bibnamefont {Amado}}, \bibinfo
  {author} {\bibfnamefont {C.}~\bibnamefont {Palomares-Garcia}}, \bibinfo
  {author} {\bibfnamefont {M.}~\bibnamefont {Alidoust}}, \bibinfo {author}
  {\bibfnamefont {K.}~\bibnamefont {Halterman}}, \bibinfo {author}
  {\bibfnamefont {M.~G.}\ \bibnamefont {Blamire}},\ and\ \bibinfo {author}
  {\bibfnamefont {J.~W.~A.}\ \bibnamefont {Robinson}},\ }\href
  {https://doi.org/10.1103/PhysRevApplied.8.044008} {\bibfield  {journal}
  {\bibinfo  {journal} {Phys. Rev. Applied}\ }\textbf {\bibinfo {volume} {8}},\
  \bibinfo {pages} {044008} (\bibinfo {year} {2017})}\BibitemShut {NoStop}%
\bibitem [{\citenamefont {Linder}\ and\ \citenamefont
  {Robinson}(2015)}]{Linder2015}%
  \BibitemOpen
  \bibfield  {author} {\bibinfo {author} {\bibfnamefont {J.}~\bibnamefont
  {Linder}}\ and\ \bibinfo {author} {\bibfnamefont {J.~W.~A.}\ \bibnamefont
  {Robinson}},\ }\href {http://dx.doi.org/10.1038/nphys3242} {\bibfield
  {journal} {\bibinfo  {journal} {Nat Phys}\ }\textbf {\bibinfo {volume}
  {11}},\ \bibinfo {pages} {307} (\bibinfo {year} {2015})}\BibitemShut
  {NoStop}%
\bibitem [{\citenamefont {Eschrig}(2015)}]{eschrig2015spin}%
  \BibitemOpen
  \bibfield  {author} {\bibinfo {author} {\bibfnamefont {M.}~\bibnamefont
  {Eschrig}},\ }\href
  {https://iopscience.iop.org/article/10.1088/0034-4885/78/10/104501}
  {\bibfield  {journal} {\bibinfo  {journal} {Reports on Progress in Physics}\
  }\textbf {\bibinfo {volume} {78}},\ \bibinfo {pages} {104501} (\bibinfo
  {year} {2015})}\BibitemShut {NoStop}%
\bibitem [{\citenamefont {Niu}(2012)}]{Niu2012}%
  \BibitemOpen
  \bibfield  {author} {\bibinfo {author} {\bibfnamefont {Z.}~\bibnamefont
  {Niu}},\ }\href {https://doi.org/10.1063/1.4743001} {\bibfield  {journal}
  {\bibinfo  {journal} {Applied Physics Letters}\ }\textbf {\bibinfo {volume}
  {101}},\ \bibinfo {pages} {062601} (\bibinfo {year} {2012})}\BibitemShut
  {NoStop}%
\bibitem [{\citenamefont {Bergeret}\ and\ \citenamefont
  {Tokatly}(2013)}]{Bergeret2013}%
  \BibitemOpen
  \bibfield  {author} {\bibinfo {author} {\bibfnamefont {F.~S.}\ \bibnamefont
  {Bergeret}}\ and\ \bibinfo {author} {\bibfnamefont {I.~V.}\ \bibnamefont
  {Tokatly}},\ }\href {https://doi.org/10.1103/PhysRevLett.110.117003}
  {\bibfield  {journal} {\bibinfo  {journal} {Phys. Rev. Lett.}\ }\textbf
  {\bibinfo {volume} {110}},\ \bibinfo {pages} {117003} (\bibinfo {year}
  {2013})}\BibitemShut {NoStop}%
\bibitem [{\citenamefont {Bergeret}\ and\ \citenamefont
  {Tokatly}(2014)}]{Bergeret2014}%
  \BibitemOpen
  \bibfield  {author} {\bibinfo {author} {\bibfnamefont {F.~S.}\ \bibnamefont
  {Bergeret}}\ and\ \bibinfo {author} {\bibfnamefont {I.~V.}\ \bibnamefont
  {Tokatly}},\ }\href {https://doi.org/10.1103/PhysRevB.89.134517} {\bibfield
  {journal} {\bibinfo  {journal} {Phys. Rev. B}\ }\textbf {\bibinfo {volume}
  {89}},\ \bibinfo {pages} {134517} (\bibinfo {year} {2014})}\BibitemShut
  {NoStop}%
\bibitem [{\citenamefont {Jacobsen}\ \emph {et~al.}(2015)\citenamefont
  {Jacobsen}, \citenamefont {Ouassou},\ and\ \citenamefont
  {Linder}}]{Jacobsen2015}%
  \BibitemOpen
  \bibfield  {author} {\bibinfo {author} {\bibfnamefont {S.~H.}\ \bibnamefont
  {Jacobsen}}, \bibinfo {author} {\bibfnamefont {J.~A.}\ \bibnamefont
  {Ouassou}},\ and\ \bibinfo {author} {\bibfnamefont {J.}~\bibnamefont
  {Linder}},\ }\href {https://doi.org/10.1103/PhysRevB.92.024510} {\bibfield
  {journal} {\bibinfo  {journal} {Phys. Rev. B}\ }\textbf {\bibinfo {volume}
  {92}},\ \bibinfo {pages} {024510} (\bibinfo {year} {2015})}\BibitemShut
  {NoStop}%
\bibitem [{\citenamefont {Silaev}\ \emph {et~al.}(2020)\citenamefont {Silaev},
  \citenamefont {Bobkova},\ and\ \citenamefont {Bobkov}}]{silaev2020odd}%
  \BibitemOpen
  \bibfield  {author} {\bibinfo {author} {\bibfnamefont {M.~A.}\ \bibnamefont
  {Silaev}}, \bibinfo {author} {\bibfnamefont {I.~V.}\ \bibnamefont
  {Bobkova}},\ and\ \bibinfo {author} {\bibfnamefont {A.~M.}\ \bibnamefont
  {Bobkov}},\ }\href
  {https://journals.aps.org/prb/abstract/10.1103/PhysRevB.102.100507}
  {\bibfield  {journal} {\bibinfo  {journal} {Physical Review B}\ }\textbf
  {\bibinfo {volume} {102}},\ \bibinfo {pages} {100507(R)} (\bibinfo {year}
  {2020})}\BibitemShut {NoStop}%
\bibitem [{\citenamefont {Bobkova}\ \emph {et~al.}(2021)\citenamefont
  {Bobkova}, \citenamefont {Bobkov},\ and\ \citenamefont
  {Silaev}}]{Bobkova2021}%
  \BibitemOpen
  \bibfield  {author} {\bibinfo {author} {\bibfnamefont {I.~V.}\ \bibnamefont
  {Bobkova}}, \bibinfo {author} {\bibfnamefont {A.~M.}\ \bibnamefont
  {Bobkov}},\ and\ \bibinfo {author} {\bibfnamefont {M.~A.}\ \bibnamefont
  {Silaev}},\ }\href {https://doi.org/10.1103/PhysRevLett.127.147701}
  {\bibfield  {journal} {\bibinfo  {journal} {Phys. Rev. Lett.}\ }\textbf
  {\bibinfo {volume} {127}},\ \bibinfo {pages} {147701} (\bibinfo {year}
  {2021})}\BibitemShut {NoStop}%
\bibitem [{\citenamefont {Jeon}\ \emph {et~al.}(2018)\citenamefont {Jeon},
  \citenamefont {Ciccarelli}, \citenamefont {Kurebayashi}, \citenamefont
  {Wunderlich}, \citenamefont {Cohen}, \citenamefont {Komori}, \citenamefont
  {Robinson},\ and\ \citenamefont {Blamire}}]{jeon2018spin}%
  \BibitemOpen
  \bibfield  {author} {\bibinfo {author} {\bibfnamefont {K.-R.}\ \bibnamefont
  {Jeon}}, \bibinfo {author} {\bibfnamefont {C.}~\bibnamefont {Ciccarelli}},
  \bibinfo {author} {\bibfnamefont {H.}~\bibnamefont {Kurebayashi}}, \bibinfo
  {author} {\bibfnamefont {J.}~\bibnamefont {Wunderlich}}, \bibinfo {author}
  {\bibfnamefont {L.~F.}\ \bibnamefont {Cohen}}, \bibinfo {author}
  {\bibfnamefont {S.}~\bibnamefont {Komori}}, \bibinfo {author} {\bibfnamefont
  {J.~W.~A.}\ \bibnamefont {Robinson}},\ and\ \bibinfo {author} {\bibfnamefont
  {M.~G.}\ \bibnamefont {Blamire}},\ }\href
  {https://journals.aps.org/prapplied/abstract/10.1103/PhysRevApplied.10.014029}
  {\bibfield  {journal} {\bibinfo  {journal} {Physical Review Applied}\
  }\textbf {\bibinfo {volume} {10}},\ \bibinfo {pages} {014029} (\bibinfo
  {year} {2018})}\BibitemShut {NoStop}%
\bibitem [{\citenamefont {Wakamura}\ \emph {et~al.}(2015)\citenamefont
  {Wakamura}, \citenamefont {Akaike}, \citenamefont {Omori}, \citenamefont
  {Niimi}, \citenamefont {Takahashi}, \citenamefont {Fujimaki}, \citenamefont
  {Maekawa},\ and\ \citenamefont {Otani}}]{wakamura2015quasiparticle}%
  \BibitemOpen
  \bibfield  {author} {\bibinfo {author} {\bibfnamefont {T.}~\bibnamefont
  {Wakamura}}, \bibinfo {author} {\bibfnamefont {H.}~\bibnamefont {Akaike}},
  \bibinfo {author} {\bibfnamefont {Y.}~\bibnamefont {Omori}}, \bibinfo
  {author} {\bibfnamefont {Y.}~\bibnamefont {Niimi}}, \bibinfo {author}
  {\bibfnamefont {S.}~\bibnamefont {Takahashi}}, \bibinfo {author}
  {\bibfnamefont {A.}~\bibnamefont {Fujimaki}}, \bibinfo {author}
  {\bibfnamefont {S.}~\bibnamefont {Maekawa}},\ and\ \bibinfo {author}
  {\bibfnamefont {Y.}~\bibnamefont {Otani}},\ }\href
  {https://doi.org/10.1038/nmat4276} {\bibfield  {journal} {\bibinfo  {journal}
  {Nature materials}\ }\textbf {\bibinfo {volume} {14}},\ \bibinfo {pages}
  {675} (\bibinfo {year} {2015})}\BibitemShut {NoStop}%
\bibitem [{\citenamefont {Wang}\ \emph {et~al.}(2017)\citenamefont {Wang},
  \citenamefont {Wang}, \citenamefont {Xie}, \citenamefont {Warsi},
  \citenamefont {Wu}, \citenamefont {Chen}, \citenamefont {Lorenz},
  \citenamefont {Fan},\ and\ \citenamefont {Xiao}}]{wang2017large}%
  \BibitemOpen
  \bibfield  {author} {\bibinfo {author} {\bibfnamefont {T.}~\bibnamefont
  {Wang}}, \bibinfo {author} {\bibfnamefont {W.}~\bibnamefont {Wang}}, \bibinfo
  {author} {\bibfnamefont {Y.}~\bibnamefont {Xie}}, \bibinfo {author}
  {\bibfnamefont {M.}~\bibnamefont {Warsi}}, \bibinfo {author} {\bibfnamefont
  {J.}~\bibnamefont {Wu}}, \bibinfo {author} {\bibfnamefont {Y.}~\bibnamefont
  {Chen}}, \bibinfo {author} {\bibfnamefont {V.}~\bibnamefont {Lorenz}},
  \bibinfo {author} {\bibfnamefont {X.}~\bibnamefont {Fan}},\ and\ \bibinfo
  {author} {\bibfnamefont {J.~Q.}\ \bibnamefont {Xiao}},\ }\href
  {https://www.nature.com/articles/s41598-017-01112-9} {\bibfield  {journal}
  {\bibinfo  {journal} {Scientific reports}\ }\textbf {\bibinfo {volume} {7}},\
  \bibinfo {pages} {1} (\bibinfo {year} {2017})}\BibitemShut {NoStop}%
\bibitem [{\citenamefont {Niimi}\ \emph {et~al.}(2011)\citenamefont {Niimi},
  \citenamefont {Morota}, \citenamefont {Wei}, \citenamefont {Deranlot},
  \citenamefont {Basletic}, \citenamefont {Hamzic}, \citenamefont {Fert},\ and\
  \citenamefont {Otani}}]{niimi2011extrinsic}%
  \BibitemOpen
  \bibfield  {author} {\bibinfo {author} {\bibfnamefont {Y.}~\bibnamefont
  {Niimi}}, \bibinfo {author} {\bibfnamefont {M.}~\bibnamefont {Morota}},
  \bibinfo {author} {\bibfnamefont {D.~H.}\ \bibnamefont {Wei}}, \bibinfo
  {author} {\bibfnamefont {C.}~\bibnamefont {Deranlot}}, \bibinfo {author}
  {\bibfnamefont {M.}~\bibnamefont {Basletic}}, \bibinfo {author}
  {\bibfnamefont {A.}~\bibnamefont {Hamzic}}, \bibinfo {author} {\bibfnamefont
  {A.}~\bibnamefont {Fert}},\ and\ \bibinfo {author} {\bibfnamefont
  {Y.}~\bibnamefont {Otani}},\ }\href
  {https://doi.org/10.1103/PhysRevLett.106.126601} {\bibfield  {journal}
  {\bibinfo  {journal} {Physical review letters}\ }\textbf {\bibinfo {volume}
  {106}},\ \bibinfo {pages} {126601} (\bibinfo {year} {2011})}\BibitemShut
  {NoStop}%
\bibitem [{\citenamefont {Ramaswamy}\ \emph {et~al.}(2017)\citenamefont
  {Ramaswamy}, \citenamefont {Wang}, \citenamefont {Elyasi}, \citenamefont
  {Motapothula}, \citenamefont {Venkatesan}, \citenamefont {Qiu},\ and\
  \citenamefont {Yang}}]{ramaswamy2017extrinsic}%
  \BibitemOpen
  \bibfield  {author} {\bibinfo {author} {\bibfnamefont {R.}~\bibnamefont
  {Ramaswamy}}, \bibinfo {author} {\bibfnamefont {Y.}~\bibnamefont {Wang}},
  \bibinfo {author} {\bibfnamefont {M.}~\bibnamefont {Elyasi}}, \bibinfo
  {author} {\bibfnamefont {M.}~\bibnamefont {Motapothula}}, \bibinfo {author}
  {\bibfnamefont {T.}~\bibnamefont {Venkatesan}}, \bibinfo {author}
  {\bibfnamefont {X.}~\bibnamefont {Qiu}},\ and\ \bibinfo {author}
  {\bibfnamefont {H.}~\bibnamefont {Yang}},\ }\href
  {https://doi.org/10.1103/PhysRevApplied.8.024034} {\bibfield  {journal}
  {\bibinfo  {journal} {Physical Review Applied}\ }\textbf {\bibinfo {volume}
  {8}},\ \bibinfo {pages} {024034} (\bibinfo {year} {2017})}\BibitemShut
  {NoStop}%
\bibitem [{\citenamefont {Bergeret}\ and\ \citenamefont
  {Tokatly}(2016)}]{bergeret2016manifestation}%
  \BibitemOpen
  \bibfield  {author} {\bibinfo {author} {\bibfnamefont {F.~S.}\ \bibnamefont
  {Bergeret}}\ and\ \bibinfo {author} {\bibfnamefont {I.~V.}\ \bibnamefont
  {Tokatly}},\ }\href
  {https://journals.aps.org/prb/pdf/10.1103/PhysRevB.94.180502} {\bibfield
  {journal} {\bibinfo  {journal} {Physical Review B}\ }\textbf {\bibinfo
  {volume} {94}},\ \bibinfo {pages} {180502(R)} (\bibinfo {year}
  {2016})}\BibitemShut {NoStop}%
\bibitem [{\citenamefont {Espedal}\ \emph {et~al.}(2017)\citenamefont
  {Espedal}, \citenamefont {Lange}, \citenamefont {Sadjina}, \citenamefont
  {Malshukov},\ and\ \citenamefont {Brataas}}]{espedal2017spin}%
  \BibitemOpen
  \bibfield  {author} {\bibinfo {author} {\bibfnamefont {C.}~\bibnamefont
  {Espedal}}, \bibinfo {author} {\bibfnamefont {P.}~\bibnamefont {Lange}},
  \bibinfo {author} {\bibfnamefont {S.}~\bibnamefont {Sadjina}}, \bibinfo
  {author} {\bibfnamefont {A.~G.}\ \bibnamefont {Malshukov}},\ and\ \bibinfo
  {author} {\bibfnamefont {A.}~\bibnamefont {Brataas}},\ }\href
  {https://journals.aps.org/prb/abstract/10.1103/PhysRevB.95.054509} {\bibfield
   {journal} {\bibinfo  {journal} {Physical Review B}\ }\textbf {\bibinfo
  {volume} {95}},\ \bibinfo {pages} {054509} (\bibinfo {year}
  {2017})}\BibitemShut {NoStop}%
\bibitem [{\citenamefont {Huang}\ \emph {et~al.}(2018)\citenamefont {Huang},
  \citenamefont {Tokatly},\ and\ \citenamefont
  {Bergeret}}]{huang2018extrinsic}%
  \BibitemOpen
  \bibfield  {author} {\bibinfo {author} {\bibfnamefont {C.}~\bibnamefont
  {Huang}}, \bibinfo {author} {\bibfnamefont {I.~V.}\ \bibnamefont {Tokatly}},\
  and\ \bibinfo {author} {\bibfnamefont {F.~S.}\ \bibnamefont {Bergeret}},\
  }\href {https://journals.aps.org/prb/abstract/10.1103/PhysRevB.98.144515}
  {\bibfield  {journal} {\bibinfo  {journal} {Physical Review B}\ }\textbf
  {\bibinfo {volume} {98}},\ \bibinfo {pages} {144515} (\bibinfo {year}
  {2018})}\BibitemShut {NoStop}%
\bibitem [{\citenamefont {Virtanen}\ \emph {et~al.}(2021)\citenamefont
  {Virtanen}, \citenamefont {Bergeret},\ and\ \citenamefont
  {Tokatly}}]{virtanen2021magnetoelectric}%
  \BibitemOpen
  \bibfield  {author} {\bibinfo {author} {\bibfnamefont {P.}~\bibnamefont
  {Virtanen}}, \bibinfo {author} {\bibfnamefont {F.~S.}\ \bibnamefont
  {Bergeret}},\ and\ \bibinfo {author} {\bibfnamefont {I.~V.}\ \bibnamefont
  {Tokatly}},\ }\href
  {https://journals.aps.org/prb/abstract/10.1103/PhysRevB.104.064515}
  {\bibfield  {journal} {\bibinfo  {journal} {Physical Review B}\ }\textbf
  {\bibinfo {volume} {104}},\ \bibinfo {pages} {064515} (\bibinfo {year}
  {2021})}\BibitemShut {NoStop}%
\bibitem [{\citenamefont {Sinova}\ \emph {et~al.}(2015)\citenamefont {Sinova},
  \citenamefont {Valenzuela}, \citenamefont {Wunderlich}, \citenamefont
  {Back},\ and\ \citenamefont {Jungwirth}}]{Sinova2015}%
  \BibitemOpen
  \bibfield  {author} {\bibinfo {author} {\bibfnamefont {J.}~\bibnamefont
  {Sinova}}, \bibinfo {author} {\bibfnamefont {S.~O.}\ \bibnamefont
  {Valenzuela}}, \bibinfo {author} {\bibfnamefont {J.}~\bibnamefont
  {Wunderlich}}, \bibinfo {author} {\bibfnamefont {C.~H.}\ \bibnamefont
  {Back}},\ and\ \bibinfo {author} {\bibfnamefont {T.}~\bibnamefont
  {Jungwirth}},\ }\href {https://doi.org/10.1103/RevModPhys.87.1213} {\bibfield
   {journal} {\bibinfo  {journal} {Rev. Mod. Phys.}\ }\textbf {\bibinfo
  {volume} {87}},\ \bibinfo {pages} {1213} (\bibinfo {year}
  {2015})}\BibitemShut {NoStop}%
\bibitem [{\citenamefont {Lifshits}\ and\ \citenamefont
  {Dyakonov}(2009)}]{Lifshits2009swapping}%
  \BibitemOpen
  \bibfield  {author} {\bibinfo {author} {\bibfnamefont {M.~B.}\ \bibnamefont
  {Lifshits}}\ and\ \bibinfo {author} {\bibfnamefont {M.~I.}\ \bibnamefont
  {Dyakonov}},\ }\href
  {https://journals.aps.org/prl/abstract/10.1103/PhysRevLett.103.186601}
  {\bibfield  {journal} {\bibinfo  {journal} {Physical review letters}\
  }\textbf {\bibinfo {volume} {103}},\ \bibinfo {pages} {186601} (\bibinfo
  {year} {2009})}\BibitemShut {NoStop}%
\bibitem [{\citenamefont {Golubov}\ \emph {et~al.}(2004)\citenamefont
  {Golubov}, \citenamefont {Kupriyanov},\ and\ \citenamefont
  {Il’Ichev}}]{golubov2004current}%
  \BibitemOpen
  \bibfield  {author} {\bibinfo {author} {\bibfnamefont {A.~A.}\ \bibnamefont
  {Golubov}}, \bibinfo {author} {\bibfnamefont {M.~Y.}\ \bibnamefont
  {Kupriyanov}},\ and\ \bibinfo {author} {\bibfnamefont {E.}~\bibnamefont
  {Il’Ichev}},\ }\href
  {https://journals.aps.org/rmp/abstract/10.1103/RevModPhys.76.411} {\bibfield
  {journal} {\bibinfo  {journal} {Reviews of modern physics}\ }\textbf
  {\bibinfo {volume} {76}},\ \bibinfo {pages} {411} (\bibinfo {year}
  {2004})}\BibitemShut {NoStop}%
\bibitem [{\citenamefont {Hoffmann}(2013)}]{hoffmann2013spin}%
  \BibitemOpen
  \bibfield  {author} {\bibinfo {author} {\bibfnamefont {A.}~\bibnamefont
  {Hoffmann}},\ }\href {https://ieeexplore.ieee.org/abstract/document/6516040}
  {\bibfield  {journal} {\bibinfo  {journal} {IEEE transactions on magnetics}\
  }\textbf {\bibinfo {volume} {49}},\ \bibinfo {pages} {5172} (\bibinfo {year}
  {2013})}\BibitemShut {NoStop}%
\bibitem [{\citenamefont {Banerjee}\ \emph {et~al.}(2018)\citenamefont
  {Banerjee}, \citenamefont {Ouassou}, \citenamefont {Zhu}, \citenamefont
  {Stelmashenko}, \citenamefont {Linder},\ and\ \citenamefont
  {Blamire}}]{banerjee2018controlling}%
  \BibitemOpen
  \bibfield  {author} {\bibinfo {author} {\bibfnamefont {N.}~\bibnamefont
  {Banerjee}}, \bibinfo {author} {\bibfnamefont {J.~A.}\ \bibnamefont
  {Ouassou}}, \bibinfo {author} {\bibfnamefont {Y.}~\bibnamefont {Zhu}},
  \bibinfo {author} {\bibfnamefont {N.~A.}\ \bibnamefont {Stelmashenko}},
  \bibinfo {author} {\bibfnamefont {J.}~\bibnamefont {Linder}},\ and\ \bibinfo
  {author} {\bibfnamefont {M.~G.}\ \bibnamefont {Blamire}},\ }\href
  {https://journals.aps.org/prb/abstract/10.1103/PhysRevB.97.184521} {\bibfield
   {journal} {\bibinfo  {journal} {Physical Review B}\ }\textbf {\bibinfo
  {volume} {97}},\ \bibinfo {pages} {184521} (\bibinfo {year}
  {2018})}\BibitemShut {NoStop}%
\bibitem [{\citenamefont {Lo}\ \emph {et~al.}(2014)\citenamefont {Lo},
  \citenamefont {Lin}, \citenamefont {Wang}, \citenamefont {Lin},\ and\
  \citenamefont {Liang}}]{lo2014spin}%
  \BibitemOpen
  \bibfield  {author} {\bibinfo {author} {\bibfnamefont {S.-T.}\ \bibnamefont
  {Lo}}, \bibinfo {author} {\bibfnamefont {S.-W.}\ \bibnamefont {Lin}},
  \bibinfo {author} {\bibfnamefont {Y.-T.}\ \bibnamefont {Wang}}, \bibinfo
  {author} {\bibfnamefont {S.-D.}\ \bibnamefont {Lin}},\ and\ \bibinfo {author}
  {\bibfnamefont {C.-T.}\ \bibnamefont {Liang}},\ }\href
  {https://www.nature.com/articles/srep05438} {\bibfield  {journal} {\bibinfo
  {journal} {Scientific reports}\ }\textbf {\bibinfo {volume} {4}},\ \bibinfo
  {pages} {1} (\bibinfo {year} {2014})}\BibitemShut {NoStop}%
\bibitem [{\citenamefont {Ast}\ \emph {et~al.}(2007)\citenamefont {Ast},
  \citenamefont {Henk}, \citenamefont {Ernst}, \citenamefont {Moreschini},
  \citenamefont {Falub}, \citenamefont {Pacil{\'e}}, \citenamefont {Bruno},
  \citenamefont {Kern},\ and\ \citenamefont {Grioni}}]{ast2007giant}%
  \BibitemOpen
  \bibfield  {author} {\bibinfo {author} {\bibfnamefont {C.~R.}\ \bibnamefont
  {Ast}}, \bibinfo {author} {\bibfnamefont {J.}~\bibnamefont {Henk}}, \bibinfo
  {author} {\bibfnamefont {A.}~\bibnamefont {Ernst}}, \bibinfo {author}
  {\bibfnamefont {L.}~\bibnamefont {Moreschini}}, \bibinfo {author}
  {\bibfnamefont {M.~C.}\ \bibnamefont {Falub}}, \bibinfo {author}
  {\bibfnamefont {D.}~\bibnamefont {Pacil{\'e}}}, \bibinfo {author}
  {\bibfnamefont {P.}~\bibnamefont {Bruno}}, \bibinfo {author} {\bibfnamefont
  {K.}~\bibnamefont {Kern}},\ and\ \bibinfo {author} {\bibfnamefont
  {M.}~\bibnamefont {Grioni}},\ }\href
  {https://journals.aps.org/prl/abstract/10.1103/PhysRevLett.98.186807}
  {\bibfield  {journal} {\bibinfo  {journal} {Physical Review Letters}\
  }\textbf {\bibinfo {volume} {98}},\ \bibinfo {pages} {186807} (\bibinfo
  {year} {2007})}\BibitemShut {NoStop}%
\end{thebibliography}%

\section{Appendix}
\label{appendix}

\subsection{Details of \texorpdfstring{$\check{J}_k$}{} and \texorpdfstring{$\check{T}$}{} calculations}

It can be checked by the straightforward calculation that

\begin{equation} \label{eq:CovDerF}
	\tilde{\nabla}_k (\check{A}\check{B}) = (\tilde{\nabla}_k \check{A}) \check{B} + \check{A} (\tilde{\nabla}_k \check{B}),
\end{equation} 
where $\check{A}$ and $\check{B}$ are  matrices in spin and Nambu spaces. Exploiting this fact we see that $\check{J}_k$ and $\check{T}$ contain terms $\propto \epsilon_{kja} \tilde{\nabla}_k \tilde{\nabla}_j \check{g}$. This combination can be simplified making use of anti-symmetric properties of the Levi-Civita tensor $\epsilon_{kja}$
\begin{equation}\label{eq:WithoutB}
	\epsilon_{kja} \tilde{\nabla}_k \tilde{\nabla}_j \check{g} = \epsilon_{kja}  \left[i (\nabla_k p_{sj})\check{\tau}_3\text{, }\check{g}\right] \propto   \frac{e}{c}(\operatorname{rot}{\mathbf{A}})_a \left[\check{\tau}_3\text{, } \check{g}\right].
\end{equation}
It means that the corresponding terms can be neglected because $D e\vert\operatorname{rot}{\mathbf{A}} \vert /c  \propto h (\xi_F p_s)(\xi_F/\lambda_L) \ll h $ in the ferromagnet or $\propto T_c (\xi_S p_s)(\xi_S/\lambda_L) \ll T_c$ in the superconductor.

%Using properties (\ref{eq:CovDerF}) and (\ref{eq:WithoutB}) we can rewrite the full Usadel equation (\ref{eq:UsadelFull}) in a more simplified way
%\begin{equation}
%\begin{aligned}
%	\left[ \omega_n \check{\tau}_3 - i \check{\Delta} + i (\mathbf{h}\cdot\hat{\bm{\sigma}})\check{\tau}_3\text{, }\check{g}\right] + \tilde{\nabla}_k \left(-D \check{g}\tilde{\nabla}_k\check{g}\right) =
%	\\= -\frac{1}{8 \tau_{so}} \left[\hat{\sigma}_a \check{g} \hat{\sigma}_a\text{, }\check{g}\right] + \check{T}_s,\\
%	\check{T}_s = -\frac{D\theta}{4} \epsilon_{kja}\left[\check{g}\tilde{\nabla}_k \check{g} \tilde{\nabla}_j \check{g}\text{, }\hat{\sigma}_a \right] -\\- \frac{i D\kappa}{4}\epsilon_{kja}\left[\tilde{\nabla}_k \check{g} \tilde{\nabla}_j \check{g}\text{, }\hat{\sigma}_a\right].
%\end{aligned}
%\end{equation}
%The minus sign of the second term in $\check{T}_s$ comes from the covariant divergence of the generalized matrix current $\check{J}_k$. Now we have to calculate parts with $\epsilon_{kja} \tilde{\nabla}_k \check{g} \tilde{\nabla}_j \check{g} \hat{\sigma}_a$ and $\hat{\sigma}_a \epsilon_{kja} \tilde{\nabla}_k \check{g} \tilde{\nabla}_j \check{g}$. With the logic already used above we arrive at
%\begin{equation}
%	\epsilon_{kja} \tilde{\nabla}_k \check{g} \tilde{\nabla}_j \check{g} \hat{\sigma}_a = \left[\nabla_x \check{g}\text{, }\left[\check{\tau}_3\text{, }\check{g}\right]\right]\left(i p_{sy} \hat{\sigma}_x - ip_{sz}\hat{\sigma}_y\right)
%\end{equation}

\subsection{Details of LRTC calculations in S/F hybrids}

\subsubsection{Calculation of the anomalous Green's function at the absolutely transparent S/F interface}

Here we provide details of calculation of the singlet and triplet anomalous Green's functions $f_{s0}$ and $\bm f_{t0} = (f_{x0},f_{y0},f_{z0})^T$ at the absolutely transparent S/F interface. The second boundary condition in Eq.~(\ref{eq:boundcoundSF}) reads as 
\begin{equation} \label{eq:boundcoundPreF}
\begin{aligned}
	-D \hat{f}' + \frac{iD \theta\operatorname{sgn}{\omega_n}}{2}\left[ p_{sy}\left\{\hat{f}\text{, }\hat{\sigma}_z\right\} -  p_{sz}\left\{\hat{f}\text{, }\hat{\sigma}_y\right\}\right] -
	\\\left. - \frac{D \kappa}{2} \left[ p_{sy}\left[\hat{f}\text{, }\hat{\sigma}_z\right] - p_{sz}\left[\hat{f}\text{, }\hat{\sigma}_y\right] \right] \right\vert_{x = -0}=\left. - D \hat{f}' \right\vert_{x = +0}.
\end{aligned}
\end{equation}
In components Eq.~(\ref{eq:boundcoundPreF}) takes the  form
\begin{equation} \label{eq:boundcoundJLinSystem}
\begin{cases}
	- f'_{sL} + i \theta\operatorname{sgn}{\omega_n}\left[p_{sy} f_{z0} - p_{sz}f_{y0}\right] = -  f'_{sR},\\
	- f'_{xL} -  i  \kappa\left[p_{sy}f_{y0} + p_{sz}f_{z0}\right] = -  f'_{xR},\\
	- f'_{yL} - i \theta \operatorname{sgn}{\omega_n} p_{sz}f_{s0} + i \kappa p_{sy}f_{x0} = -  f'_{yR},\\
	- f'_{zL} + i \theta\operatorname{sgn}{\omega_n} p_{sy}f_{s0} + i  \kappa p_{sz}f_{x0} = -  f'_{zR}.
\end{cases}
\end{equation}
Here $f'_{i{(L,R)}} = \frac{df_{i}(x = \mp0)}{dx}$, $i = (s, x, y, z)$. The interface Green's function components $f_{s0}$, $f_{x0}$, $f_{y0}$, $f_{z0}$ are fully determined by the system of linear equations (\ref{eq:boundcoundJLinSystem}) with the solutions (\ref{eq:UsadelSSol}) and (\ref{eq:UsadelFSol}).  

\subsubsection{Calculation of the LRTC in S/F/S Josephson junction with absolutely transparent interfaces}

Under the condition $d_F \gg \xi_F$ the Josephson current in S/F/S Josephson junction is only carried by the LRTC. Here we provide details of the calculation of LRTC in the S/F/S junction. The continuity of the matrix current $\check J_x$ at each of the S/F interfaces results in the following system of equations:
\begin{equation} \label{eq:boundcoundJJ}
\begin{cases}
	-f'_{xSL} -  i  \kappa\left[p_{sy}f_{yL} + p_{sz}f_{zL}\right] = -  f'_{xFL},\\
	- f'_{ySL} - i  \theta \operatorname{sgn}{\omega_n} p_{sz}f_{s0L} + i \kappa p_{sy}f_{xL} = -  f'_{yFL},\\
	-  f'_{xFR} = - f'_{xSR} +  i \kappa\left[p_{sy}f_{yR} + p_{sz}f_{zR}\right],\\
	-  f'_{yFR} = - f'_{ySR} + i \theta \operatorname{sgn}{\omega_n} p_{sz}f_{sR} -i \kappa p_{sy}f_{xR}.\\
\end{cases}
\end{equation}
Here indexes $L$ and $R$ mean left ($x = 0$) and right ($x = d_F$) interfaces. We put $p_{s(z,y)L} = p_{s(z,y)} = - p_{s(z,y)R}$. We again assume $\tau_{so} \gg 1/T_c$ and  $D_S = D_F = D$ for simplicity.
The SRC $f_s$ and $f_{z}$, generated at the both S/F interfaces, do not overlap and, therefore, coincide with that ones calculated at the separate S/F interface with substitution of the appropriate $\Delta(x)$. From Eq.~(\ref{eq:FSolLinInThetaKappa_sz}) it can be obtained that \begin{equation} \label{eq:UsadelSRT}
\begin{aligned}
	f_{s(L,R)} = -\frac{i\lambda_s \xi_F (1+\lambda_s \xi_F)\vert\Delta\vert e^{\pm i\varphi/2}}{Z\vert\omega_n\vert \iffalse + \frac{D p^2_s}{2} \fi},\\
	f_{z(L,R)} =-\frac{\lambda_s \xi_F\vert\Delta\vert e^{\pm i\varphi/2}}{\omega_n \iffalse + \frac{D p^2_s}{2}\fi}.
\end{aligned}
\end{equation}  
The solutions of the Usadel equations in the $S$ layers take the form:
\begin{equation} \label{eq:UsadelSolIn2SJJ_app}
\begin{aligned}
	\begin{pmatrix}
		f_{x}\\
		f_{y}
	\end{pmatrix} = \begin{pmatrix}
		f_{xL}\\
		f_{yL}
\end{pmatrix} e^{\lambda_s x}\text{ for $x \leq 0$, and  }\\ \begin{pmatrix}
f_{x}\\
f_{y}
\end{pmatrix} = \begin{pmatrix}
f_{xR}\\
f_{yR}
\end{pmatrix} e^{-\lambda_s(x - d_F)} \text{ for $x \geq d_F$.}
\end{aligned}
\end{equation} 
In the ferromagnetic layer $0 < x < d_F$ we get
\begin{equation}\label{eq:UsadelSolInFJJ}
\begin{aligned}
	\begin{pmatrix}
		f_{x}\\
		f_{y}
	\end{pmatrix} = \frac{1}{2 \sinh{\lambda_s d_F}}\begin{pmatrix}
		f_{xL} e^{\lambda_s d_F} - f_{xR}\\
		f_{yL} e^{\lambda_s d_F} - f_{yR}\\
	\end{pmatrix} e^{-\lambda_s x} +
	\\ + \frac{1}{2 \sinh{\lambda_s d_F}}\begin{pmatrix}
	f_{xR} e^{\lambda_s d_F} - f_{xL}\\
	f_{yR} e^{\lambda_s d_F} - f_{yL}\\
\end{pmatrix} e^{\lambda_s(x - d_F)}.
\end{aligned}
\end{equation}
The continuity of the Green functions at the interfaces has already been implemented in Eqs.~(\ref{eq:UsadelSolIn2SJJ_app}) and (\ref{eq:UsadelSolInFJJ}). Combining (\ref{eq:boundcoundJJ}), (\ref{eq:UsadelSolIn2SJJ_app}), (\ref{eq:UsadelSolInFJJ})  we build a linear system for four variables $f_{xL}$, $f_{yL}$ and $f_{xR}$, $f_{yR}$.  Solving this system and substituting the quantities into Eq.~(\ref{eq:UsadelSolInFJJ}) one obtains Eq.~(\ref{eq:LRTCideal}) for the LRTC in the ferromagnetic interlayer of the Josephson junction.

\subsection{Condensate momentum distribution in S/F and S/F/S structures}

In this section we calculate the distribution of the condensate momentum induced by the external field in the superconducting parts of S/F and S/F/S structures. Our consideration follows \cite{silaev2020odd}.  In S/F bilayer we have 
\begin{equation}
	\mathbf{B}(x) = \begin{cases}
		(0\text{, } B_{0y}e^{x/\lambda_L}\text{, } B_{0z}e^{x/\lambda_L})^T \text{ for } x \leq 0,\\
		(0\text{, } B_{0y}\text{, } 4\pi M_z + B_{0z})^T \text{ for }0 <x < d_F.
	\end{cases}
\end{equation}
Here $T$ denotes transpose operation, $\lambda_L$ is the London penetration depth and $M_z$ is the appropriate component of the magnetization. By integration we come to the vector-potential
\begin{equation}
\begin{aligned}
	\mathbf{A}(x) =\\= \begin{cases}
		\begin{pmatrix}0\\ 
		{\lambda_L}B_{0z}(e^{x/\lambda_L} - 1)\\ -{\lambda_L}B_{0y}(e^{x/\lambda_L} - 1)\end{pmatrix} \text{ for } x \leq 0\\
		\begin{pmatrix}0\\ 4\pi M_z x + B_{0z} x\\ -B_{0y} x\end{pmatrix} \text{ for }0 <x < d_F.
	\end{cases}
\end{aligned}
\end{equation}
The continuity of the vector potential at the interface has already been implemented. As at $x \to -\infty$  $\mathbf{A}(x)$ does not tend to $0$, the phase gradient $\nabla \varphi = \frac{2e}{mc}(0, -\lambda_L B_{0z}, \lambda_L B_{0y})$ have to exist to compensate the condensate motion in the bulk. Therefore, the condensate momentum takes the form
\begin{equation}
\begin{aligned}
	\mathbf{p}_s = m\left[\nabla\varphi - \frac{2e}{mc}\mathbf{A}(x)\right]=\\=
		\frac{2e}{c}\begin{pmatrix}0\\ -\lambda_L B_{0z} e^{x/\lambda_L}\\ \lambda_L B_{0y} e^{x/\lambda_L}\end{pmatrix} \text{ for } x\leq 0.
\end{aligned}
\end{equation}

Analogously, for the S/F/S structure we have  \begin{equation}
	\mathbf{B}(x) = \begin{cases}
		\begin{pmatrix}0\\ B_{0y}e^{x/\lambda_L}\\ B_{0z}e^{x/\lambda_L}\end{pmatrix} \text{ for } x \leq 0\\
		\begin{pmatrix}0\\ B_{0y}\\ 4\pi M_z + B_{0z}\end{pmatrix} \text{ for }0 <x < d_F,\\
		\begin{pmatrix}0\\ B_{0y}e^{-(x - d_F)/\lambda_L}\\ B_{0z}e^{-(x - d_F)/\lambda_L}\end{pmatrix} \text{ for } x \geq d_F,
	\end{cases}
\end{equation}
%This magnetic field profile leads to the vector potential 
%\begin{equation}
%	\mathbf{A}(x) = \begin{cases}
%		\begin{pmatrix}0,\\ {\lambda_L}B_{0z}(e^{x/\lambda_L} - 1),\\ -{\lambda_L}B_{0y}(e^{x/\lambda_L} - 1)\end{pmatrix} \text{ for } x \leq 0,\\
%		\begin{pmatrix}0,\\ 4\pi M_z x + B_{0z} x,\\ -B_{0y} x\end{pmatrix} \text{ for }0 <x < d_F,\\
%		\begin{pmatrix}0,\\ ( 4\pi M_z  + B_{0z} )d_F -{\lambda_L}B_{0z}(e^{-(x - d_F)/\lambda_L} - 1),\\ -B_{0y}d_F + {\lambda_L}B_{0y}(e^{-(x - d_F)/\lambda_L} - 1))\end{pmatrix}\\
%		\text{ for }x \geq d_F.
%	\end{cases}
%\end{equation}{\color{red}

 what results in
\begin{equation} \label{eq:psInJJ}
\begin{aligned}
	\mathbf{p}_s = m\left[\nabla\varphi - \frac{2e}{mc}\mathbf{A}(x)\right]=\\ = \begin{cases}
		\frac{2e}{c}\begin{pmatrix}0\\ -\lambda_L B_{0z} e^{x/\lambda_L}\\ \lambda_L B_{0y} e^{x/\lambda_L}\end{pmatrix} \text{ for } x\leq 0,\\
		\frac{2e}{c}\begin{pmatrix}0\\ \lambda_L B_{0z} e^{-(x - d_F)/\lambda_L}\\ -\lambda_L B_{0y} e^{-(x - d_F)/\lambda_L}\end{pmatrix} \text{ for } x \geq d_F. 
	\end{cases}
\end{aligned}
\end{equation}

\end{document}